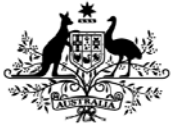

**Australian Government**

**Department of Agriculture and Water Resources**

ABARES

# A micro-simulation model of irrigation farms in the southern Murray-Darling Basin

Huong Dinh, Manannan Donoghoe, Neal Hughes, Tim Goesch[1]

[1]All authors are in The Australian Bureau of Agricultural and Resource Economics and Sciences (ABARES), Canberra, Australian Capital Territory. The views expressed in this article are the author's own and do not necessarily reflect the views of ABARES.

# Abstract

This paper presents a farm level irrigation microsimulation model of the southern Murray-Darling Basin. The model leverages detailed ABARES survey data to estimate a series of input demand and output supply equations, derived from a normalised quadratic profit function. The parameters from this estimation are then used to simulate the impact on total cost, revenue and profit of a hypothetical 30 per cent increase in the price of water. The model is still under development, with several potential improvements suggested in the conclusion. This is a working paper, provided for the purpose of receiving feedback on the analytical approach to improve future iterations of the microsimulation model.

# About

This paper presents a new farm level microsimulation model of irrigation activity within the southern Murray-Darling Basin (sMDB), developed by ABARES on behalf of the Department of Agriculture and Water Resources. The model remains under ongoing development, a number of planned improvements to the model are documented in the paper.

This paper remains technical in nature and documents the model data sources and assumptions. The results presented in this paper are hypothetical and provided for the purposes of validating the models performance. The model has a range of potential future applications, such as assessing the effects on sMDB irrigation industries of: water policy reforms, changes to climate and input and output prices.

# 1 Introduction

Irrigation farms within the southern Murray–Darling Basin (sMDB) have been subject to a wide range of climate, market and policy shocks over the last decade. Firstly, there have been dramatic variations in water availability including the Millennium drought and subsequent floods, along with a longer-term trend toward reduced winter rainfall in the region related to climate change (BOM, CSIRO 2012). In addition, changes in commodity prices and technology have transformed the irrigation sector, leading to expansions in some sectors (e.g., nuts and cotton) and contractions in others (e.g., grapes and dairy). At the same time, there have been a number of government policies associated with the Murray-Darling Basin Plan, focused on recovering water from irrigation farms for environmental uses.

As a result there is much interest in measuring and separating the effects of different climate, market and policy shocks on the profitability of irrigation farms in the sMDB. This report presents a new micro-simulation model of irrigation farms in the sMDB, developed using ABARES farm survey data. This model simulates farm input use, output supply, revenue, cost and profit, given fixed inputs (e.g., land and capital), climate conditions (e.g., rainfall), input prices (e.g., the price of water) and output prices.

The model is estimated using a sample of 3,627 irrigation farms in the sMDB between the period 2006-07 and 2014-15. The model includes four irrigation industries or farm types: dairy, horticulture, broadacre (rice) and broadacre (non-rice). For each of these industries a set of input demand and output supply functions are estimated econometrically. These functions are then combined into a micro-simulation model. Taking the 2014-15 surveyed farms as a baseline, the model simulates changes in input use, output supply and profit in response to a given exogenous shock.

The model has a range of potential applications. In particular, the model can be used to assess the effects on farm output, input and profit of changes in water market prices, due either to changes in climate or water policy. To this end, the micro-simulation model complements ABARES recently developed Water Trade Model (WTM) (Gupta et al. 2018). For example, the WTM model could be used to simulate the effects on water market prices of changes in policy (i.e., environmental water recovery), these results could then be applied to the micro-simulation model to measure effects on irrigation farm profitability. The model could equally be applied to simulate changes in other exogenous variables, such as the price of key outputs (such as rice, milk or grapes) or changes to climate (farm rainfall).

Similar to the WTM, the micro-simulation model is short-run in nature, simulating annual changes in farm production holding farm land areas and infrastructure fixed. The micro-simulation model is designed to take account of the different types of short-run farm responses observed across irrigation industries. For example, in response to water price increases dairy farms have the ability to substitute irrigation water for livestock fodder. Broadacre farms have flexibility in the short-run to vary crop areas and therefore water requirements, while horticulture farms have limited flexibility due to their permanent tree plantings.

This report begins with a description of the data. Section 3 then provides a brief overview of the model structure and the econometric estimation process (a complete treatment of the model is presented in Appendix A). Section 4 demonstrates the micro-simulation model for an arbitrary 30 per cent water price increase scenario, presenting results for average farm changes in input, output and profit in each industry. Finally Section 5 offers some conclusions and outlines some options for further research.

# 2 Data

## ABARES Murray-Darling Basin Irrigation Survey (MDBIS)

The data for this study is drawn primarily from the ABARES Murray-Darling Basin Irrigation Survey (MDBIS). The MDBIS is an annual survey of irrigated dairy, broadacre and horticulture farms throughout the MDB commencing in 2006-07. The survey includes farm-level physical and financial data, including land area and value, crop and livestock production and sales, irrigation water use, farm receipts and costs, farmer characteristics, debt and farm capital. More detail about ABARES survey methods can be found in Ashton and Oliver (2011) and Ashton et al. (2013). Note that the statistics presented throughout this report differ from those in ABARES main farm survey publications due to differences in methodology (including for example treatment of outliers and the use of survey weights).

The survey provides coverage of three industry categories: dairy, horticulture and broadacre. In this study, we further separate broadacre farms into rice and non-rice farms. Here, rice farms are those deemed to have the capacity to produce rice (based on their location and production history). Rice farms operate on an opportunistic basis and so may not necessarily plant rice crops each year.

The number of observations by year and industry are provided in figure 1 below. For this study, a total of 3, 579 farm observations were available in the sMDB between 2006-07 and 2014-15. This includes 1, 740 horticulture observations, 864 in dairy, 622 in broadacre (non-rice) and 353 farms in the broadacre (rice) industry. Within the southern Murray-Darling Basin the survey provides coverage of 6 catchment regions including: Goulburn-Broken, Loddon-Campaspe, Murrumbidgee, Murray, Lower Darling and Mt Lofty (Map 1).

**Figure 1 MDBIS observations in the sMDB per year by industry**

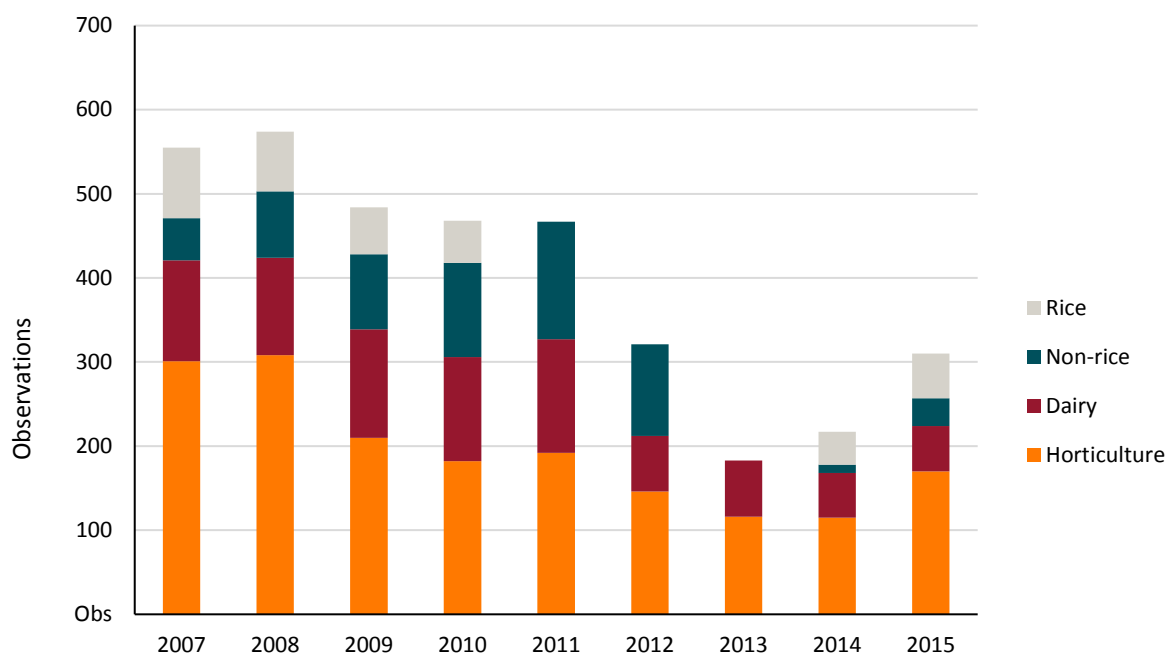

**Map 1 Microsimulation regions**

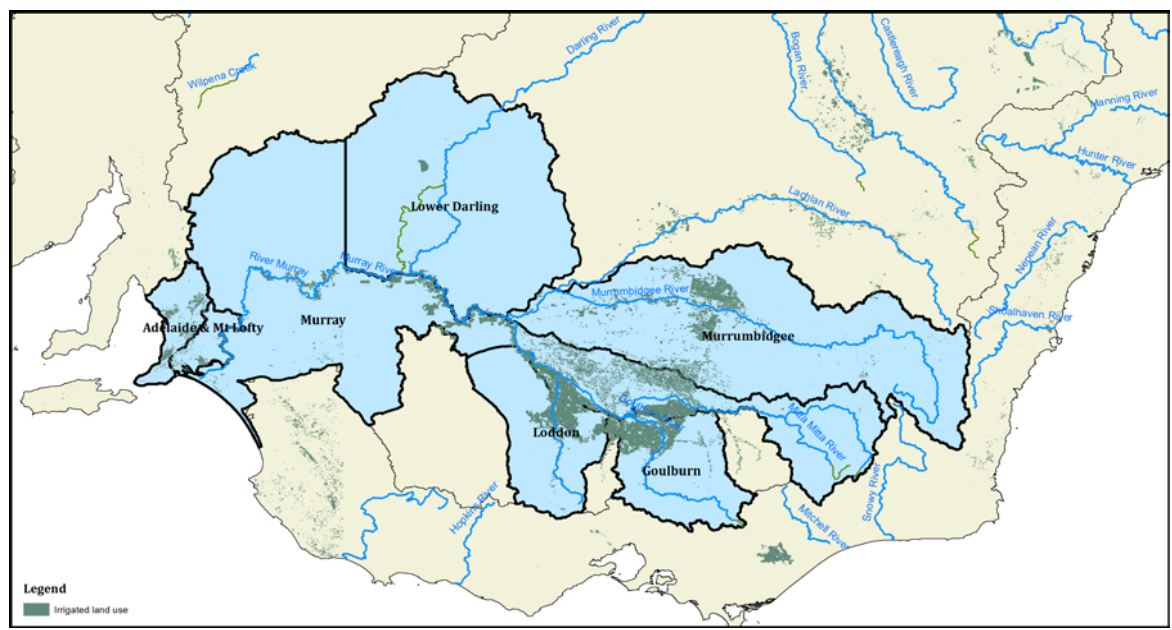

The data set used to estimate the model defines a total of 12 outputs and 4 variable inputs, along with a range of fixed inputs and other control variables. As shown in Figure 2 different sets of outputs and inputs are defined for each industry. Each of these outputs and inputs are detailed further below.

**Figure 2 Key variables in the irrigation micro-simulation model**

| Dependent variables | Independent variables |
|---|---|

| **Inputs, $x$** | **Outputs, $y$** | **Prices, $p$** |
|---|---|---|
| • Labour | **Dairy** | • Input prices |
| • Materials and services | • Milk | • Output prices |
| • Water use | • Dairy cattle | |
| • Fodder (dairy only) | **Broadacre** | |
| | • Rice (rice farms only) | **Controls, $z$** |
| | • Livestock | • Area operated |
| | • Other broadacre, incl. wheat | • Family labour |
| | **Horticulture** | • Opening livestock numbers (Beef, Sheep, Dairy) |
| | • Pome fruit | • Total capital value |
| | • Stone fruit | • Value of water entitlements |
| | • Citrus | • Average rainfall |
| | • Table grapes | • Education level |
| | • Wine grapes | • Age |
| | • Vegetables | |
| | • Other horticulture | |

# Output quantities

Table 1 provides summary statistics for outputs in each industry over the whole survey period. In the dairy industry, two outputs are defined: milk and dairy cattle. Milk output is measured as litres of milk produced. Dairy cattle output is measured as the number of cattle sold plus closing numbers minus opening numbers.

In the broadacre (rice) industry, three outputs are defined: rice, broadacre crops (including wheat) and livestock. Aside from livestock, all outputs are measured in number of tonnes produced. Livestock is measured as the number of animals sold plus the difference between the closing and opening numbers. Broadacre (non-rice) farms includes a similar set of outputs to rice farms, however rice production is excluded as an output.

In the horticulture industry, seven outputs are defined: pome fruit (for example, apples, pears and nashi fruits), stone fruit (for example, peaches, nectarines and plums), citrus fruit, table grapes, wine grapes, vegetables and other horticulture products. All horticultural outputs are measured in number of tonnes produced.

**Table 1 Summary statistics of outputs, by industry**

| Output | 2006-07 to 2014-15 | | |
|---|---|---|---|
| | **Mean** | **SD** | **Median** |
| **Dairy** | | | |
| Milk (L) | 2,001,957 | 1,733,638 | 1,530,108 |
| Dairy cattle (no.) | 264 | 205 | 215 |
| **Broadacre (rice)** | | | |
| Rice (t) | 1,158 | 960 | 950 |
| Other Broadacre, incl. wheat (t) | 1,281 | 2,495 | 589 |
| Livestock (no.) | 1,102 | 1,371 | 719 |
| **Broadacre (non-rice)** | | | |
| Other Broadacre crops (t) | 1,695 | 2,731 | 783 |
| Livestock (no.) | 1,270 | 1,562 | 758 |
| **Horticulture** | | | |
| Pome fruit (t) | 1,082 | 1,472 | 396 |
| Citrus (t) | 833 | 1,346 | 444 |
| Stone fruit (t) | 369 | 736 | 134 |
| Table grapes (t) | 312 | 682 | 120 |
| Wine grapes (t) | 1,008 | 1,820 | 306 |
| Vegetables (t) | 3,508 | 9,045 | 880 |
| Other horticulture (t) | 361 | 779 | 89 |

Note: Summary statistics are calculated at the farm level

# Variable input quantities

Table 2 provides summary statistics of variable inputs by industry. In all industries, the model includes three variable inputs: hired labour, water and materials and services (for example fertiliser, electricity and fuel). Materials and services do not include interest payments. Fodder is also included in the dairy industry to account for the potential substitution between fodder and water.

Hired labour is measured in number of weeks worked and includes both permanent and casual farm labour. Family labour is not included as a variable input but is used as a control variable. The quantity of water is measured as total megalitres of water used in production. The quantity of materials and services used is calculated by dividing total expenditure on materials and services by ABARES weighted materials and services price index. Fodder is measured by dividing fodder expenditure by ABARES price index for fodder.

**Table 2 Summary statistics of variable inputs by industry**

| Industry | Input | 2006-07 to 2014-15 | | |
|---|---|---|---|---|
| | | **Mean** | **SD** | **Median** |
| Dairy | Fodder quantity index | 352,380 | 386,143 | 249,713 |
| | Hired labour (weeks) | 100 | 131 | 70 |
| | Water (ML) | 530 | 600 | 370 |
| | Materials and services index | 796,116 | 674,395 | 635,635 |
| Broadacre (rice) | Hired labour (weeks) | 78 | 101 | 48 |
| | Water (ML) | 1,363 | 1,733 | 710 |
| | Materials and services index | 542,933 | 576,681 | 359,138 |
| Broadacre (non-rice) | Hired labour (weeks) | 80 | 134 | 52 |
| | Water (ML) | 1,013 | 2,091 | 270 |
| | Materials and services index | 496,239 | 714,552 | 281,019 |
| Horticulture | Hired labour (weeks) | 265 | 534 | 84 |
| | Water (ML) | 488 | 1,114 | 150 |
| | Materials and services index | 704,020 | 1,540,585 | 232,735 |

Note: Summary statistics are calculated at the farm level

# Fixed inputs and exogenous variables

The micro-simulation model also includes other independent variables that can affect input demand and output supply, shown in Table 3. That is inputs which are assumed to be fixed in the short run.

## Fixed inputs

Following Arnade and Kelch (2007), this study includes  family labour, land, capital and opening livestock numbers as fixed inputs.

Family labour is measured as the number of family members working on the farm.

Land is measured as the total opening number of hectares under operation for a given financial year. In the horticulture industry, the area of land allocated to each output is fixed in the medium term so ABARES has used area of land operated for each output in the output equations.

Capital is measured as the total opening value of vehicles, machinery and equipment used on the farm. Farms where a zero value has been recorded for capital are assumed to be data collection errors. These observations are replaced with an imputed value rather than being removed. This is to maintain as much of the initial sample as possible, as the farm production model requires a relatively large sample size to work effectively. The imputed value is calculated as the difference between total farm capital (which includes land, water entitlements and other capital) and the value of land and water entitlements.

Sheep, beef and dairy cattle are measured in numbers of animals at the beginning of the financial year.

## Exogenous variables

The data set contains a range of control variables, including water entitlement holdings, rainfall, region, education and age. The same control variables are applied to all industries, unless otherwise indicated.

Water entitlements are measured as the market value of water entitlements at the beginning of the financial year.

Rainfall is measured as total millilitres of annual rainfall on farm. This data is obtained by mapping the location of farms to spatial rainfall data from the Australian Water Availability Project (AWAP) (csiro.au/awap).

Education identifies if the farm operator has completed a tertiary education qualification at the time of the survey. Age is the age of the farm operator at the time of the survey.

Region is defined as the geographical catchment area where a farm is located as shown in (Map 1).

Table 3 Summary statistics of fixed inputs and exogenous variables by industry

| Industry | Fixed and Exogenous variables | 2006-07 to 2014-15 | | | Baseline (2014-15) | | |
|---|---|---|---|---|---|---|---|
| | | Mean | SD | Median | Mean | SD | Median |
| **Dairy** | Area operated (ha) | 357 | 359 | 238 | 518 | 347 | 398 |
| | Family labour (no.) | 3 | 1 | 3 | 3 | 2 | 2 |
| | Dairy cattle opening number (no.) | 436 | 356 | 332 | 710 | 499 | 569 |
| | Total capital value ($) | 565,512 | 524,945 | 420,840 | 944,343 | 906,527 | 728,052 |
| | Value of water entitlements ($) | 1,068,682 | 1,106,846 | 824,990 | 1,375,966 | 2,154,614 | 1,054,560 |
| | Average rainfall (ml) | 455 | 220 | 365 | 349 | 162 | 295 |
| | University educated (0/1) | 0.1 | 0.3 | 0 | 0.2 | 0.4 | 0 |
| | Age | 55 | 11 | 55 | 55 | 11 | 54 |
| **Broadacre (Rice)** | Area operated (ha) | | | | | | |
| | | 1,343 | 2,626 | 669 | 1,489 | 1,702 | 975 |
| | Family labour (no.) | 3 | 1 | 2 | 2 | 1 | 2 |
| | Beef cattle opening number (no.) | 77 | 151 | 0 | 67 | 102 | 4 |
| | Sheep opening number (no.) | 1,019 | 1,760 | 498 | 1,011 | 1,425 | 500 |
| | Total capital value ($) | 617,277 | 634,487 | 381,132 | 807,695 | 913,311 | 381,132 |
| | Water entitlement value ($) | 1,864,854 | 1,962,388 | 1,306,127 | 1,863,217 | 1,643,771 | 1,296,957 |
| | Average rainfall (ml) | 339 | 80 | 345 | 342 | 70 | 357 |
| | University educated (0/1) | 0.2 | 0.4 | 0 | 0.3 | 0.4 | 0 |
| | Age | 57 | 11 | 57 | 60 | 12 | 60 |
| **Broadacre (Non-Rice)** | Area operated (ha) | | | | | | |
| | | 1,932 | 3,411 | 800 | 2,700 | 3,943 | 728 |
| | Family labour (no.) | 3 | 2 | 2 | 3 | 1 | 2 |
| | Beef cattle opening number (no.) | 79 | 339 | 0 | 268 | 528 | 120 |
| | Sheep opening number (no.) | 1,288 | 1,926 | 563 | 1,803 | 3,612 | 822 |
| | Total capital value ($) | 605,719 | 662,762 | 351,904 | 872,286 | 1,275,786 | 494,698 |
| | Water entitlement value ($) | 1,246,499 | 3,554,448 | 811,330 | 860,340 | 908,466 | 462,384 |
| | Average rainfall (ml) | 522 | 171 | 498 | 396 | 134 | 353 |
| | University educated (0/1) | 0.2 | 0.4 | 0 | 0.2 | 0.4 | 0 |
| | Age | 58 | 12 | 57 | 59 | 8 | 60 |
| **Horticulture** | Area operated (ha) | 123 | 603 | 20 | 119 | 836 | 20 |
| | Family labour (no.) | 2 | 1 | 2 | 2 | 1 | 2 |
| | Beef cattle opening number (no.) | 5 | 46 | 0 | 1 | 13 | 0 |
| | Sheep opening number (no.) | 25 | 254 | 0 | 28 | 272 | 0 |
| | Capital value of vehicles ($) | 121,154 | 209,537 | 68,563 | 111,610 | 141,091 | 76,050 |
| | Other capital value ($) | 117,469 | 307,420 | 46,615 | 112,056 | 190,923 | 61,854 |
| | Water entitlements value ($) | 591,804 | 1,250,984 | 274,253 | 420,806 | 624,531 | 206,248 |
| | Rainfall (ml) | 342 | 197 | 291 | 274 | 93 | 259 |
| | University educated (0/1) | 0.2 | 0.4 | 0 | 0.2 | 0.4 | 0 |
| | Age | 58 | 10 | 59 | 60 | 10 | 62 |
| | Pome fruit area (ha) | 30 | 40 | 12 | 28 | 38 | 12 |
| | Citrus area (ha) | 35 | 53 | 18 | 37 | 37 | 21 |
| | Stone fruit area (ha) | 23 | 44 | 10 | 19 | 40 | 7 |
| | Tabletop grape area (ha) | 30 | 56 | 11 | 23 | 25 | 17 |
| | Wine grape area (ha) | 64 | 105 | 25 | 69 | 102 | 30 |
| | Vegetable area (ha) | 85 | 199 | 28 | 88 | 117 | 48 |
| | Other horticulture area (ha) | 60 | 123 | 12 | 18 | 20 | 15 |

Note: Summary statistics are calculated at the farm level and are weighted

## Output prices

The microsimulation model uses prices for inputs and outputs for the period 2006-07-2014-15. Commodity prices used in this analysis are unit values derived by dividing annual gross value data for each commodity (ABS Cat. 7503) by volume of production data for that commodity (ABS

Cat. 7121). All prices are adjusted by the CPI, and are measured in 2014-15 dollar values or in an index of 2014-15 prices. Output prices are assumed to be uniform across regions. For the broadacre (non-rice) industry, the wheat price is used as the price for other broadacre crops.

## Variable input prices

Figure 3 gives the water allocation prices from 2006-07 to 2014-15, for each region included in the model. Water market prices are taken from Gupta et al. (2018) who collated annual average prices of water allocations from a variety of sources, including state water registers, water exchanges and published reports. For the farm production and microsimulation model, Adelaide & Mt Lofty, NSW Murray and Victorian Murray are all assumed to have the same water price as do the Goulburn and Loddon regions.

**Figure 3 Mean water allocation prices by region, 2006-07 to 2014-15**

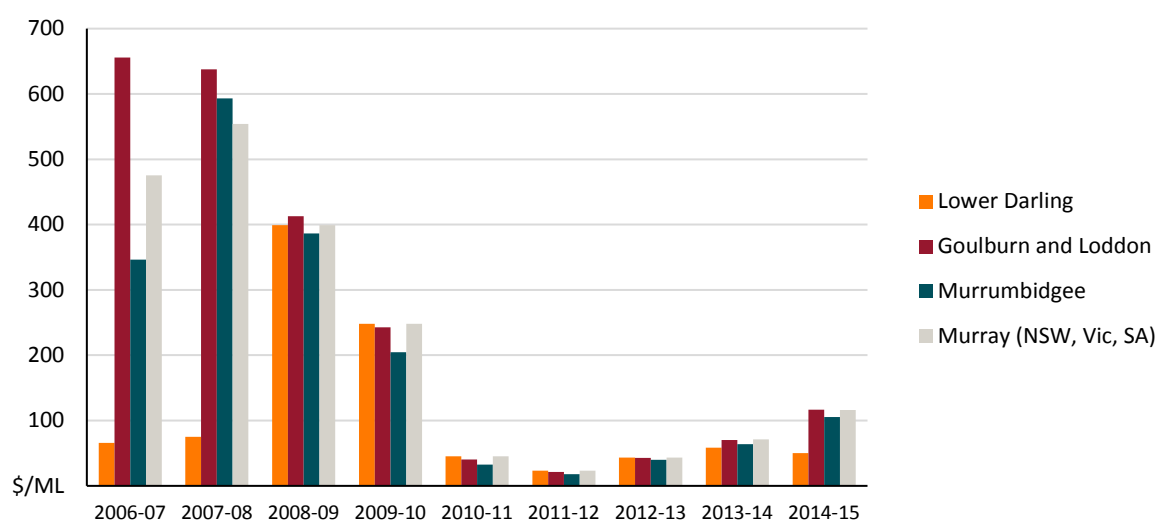

The hired labour price is calculated as the ratio of wages paid to the quantity of hired labour in weeks using ABARES MDBIS data. The median price of labour is used as the predicted price because the distribution of the wage rate is highly skewed.

Other input price indices are sourced from a combination of published data, including data published by the ABS and ABARES. The list of indices used includes: fodder; fuel; electricity; seed; chemicals and fertiliser. ABARES has previously published these estimates for Meat & Livestock Australia's 'Beef Producer Price Index', and they underlie ABARES estimates of total factor productivity in agriculture. The fodder index is only used for the dairy industry. For all industries, the price index for materials and services is calculated as the expenditure-weighted average of the price indices for fertiliser, electricity, chemicals, fuel, seed and other non-water materials and services.

# 3 Model structure and estimation

## Model structure

The irrigation farm micro-simulation model involves a series of input demand / output supply (or 'netput') functions. Each of these functions predicts the use of a specific input (supply of an output) as a function of input and output prices, fixed inputs and other control variables. Each of the four defined industries (dairy, horticulture, broadacre-rice and, broadacre-non-rice) are modelled separately.

A basic outline of the theoretical framework of the model is provided in figure 4 below, and described further in Appendix A. This framework assumes that each farm input and output choices are made to maximize short-run (annual) profit, taking fixed inputs (such as land and capital) as given. More specifically the approach derives the netput functions from a standard 'normalised quadratic profit function'.

In this study, the profit function (and resulting netput functions) are specified on a 'per hectare' basis for the horticulture and the dairy. Output and input variables are divided by total area operated for the dairy industry and by total horticultural area for the horticultural industry. This approach helps take account of the important effects of farm size on input and output responses.

Broadacre farms (rice and non-rice) are not modelled on a per hectare basis. With broadacre farms area operated is a relatively poor indicator of farm scale, given that land use intensity (e.g., proportion of land area cropped / irrigated) can vary dramatically between farms. For this reason, modelling aggregate, rather than per hectare outputs in the broadacre industry tends to improve the fit of the model.

Figure 4 Farm production and micro-simulation model setup

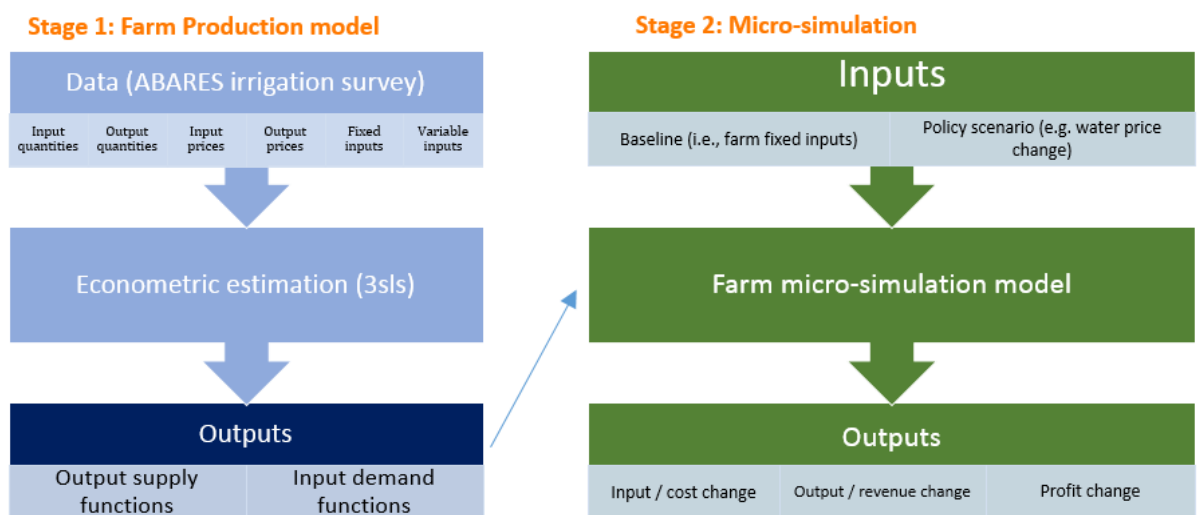

## Estimation

For each industry type, the system of netput functions is estimated via three stage least squares regression (Woolridge 2012). Further detail on the econometric estimation can be found in appendix A. The full set of parameter estimates and standard errors are presented in appendix B, while the full set of own and cross-price elasticities are presented in appendix C.

## Water price elasticities

Table 4 below, presents the own price elasticities of water for each industry, these figures estimate the average change in water demand across all farms to a 1 per cent increase in the price of water.

The own price elasticities for water show that the demand for water on broadacre (rice) farms is more elastic (more sensitive to a change in price) than those in the dairy industry, which are more elastic than horticulture farms. For example, all else held constant a 10 per cent increase in water price leads to an average 22.3 per cent reduction in water use on rice farms, compared with a 11.8 per cent for non-rice broadacre farms, 4.9 per cent on dairy farms and a close to 0 per cent reduction on horticulture farms. While the own price elasticity for horticultural farms is positive, the p value is insignificant, meaning the model essentially estimates an inelastic production response in the short run.

This is expected given that broadacre farms have much more flexibility to vary production and crop areas on an annual basis, while horticulture farms have little capacity to reduce water demand in the short-run, given the need to maintain permanent plantings.

Table 4 Own price elasticity of water by industry

| Industry | Own price elasticity of water | p-value |
|---|---|---|
| Dairy | -0.49 | 0 |
| Broadacre (rice) | -2.23 | 0 |
| Broadacre (non-rice) | -1.18 | 0 |
| Horticulture | 0.01 | 0.9 |

## Water demand curves

The estimated parameters can be used to construct water demand curves, which show volume of water used by a farm for a range of possible water prices (Figure 5) holding all other factors constant (at their mean values). In each industry, a unique water demand curve is calculated for four quartiles of area operated – quartile one representing the smallest farms and quartile four, the largest. The point where water use is zero shows the model's estimate of the maximum market price at which farms in each industry are willing to purchase water. In practice, this curve (and the maximum prices) will shift up or down depending on other factors, for example it will typically shift up when output prices are up.

Broadacre (rice-farms) are the most responsive to water price increases. Dairy is also relatively responsive, ceasing to purchase water between prices between $900 and $700 per ML depending on the farm size. Horticulture farms are most unresponsive to water price increases, with the model unable to pick up a demand response under the short period of data used in the model. For this reason, the horticulture demand curves shown are perfectly inelastic.

Given the 'per hectare' functional form, a unique water demand curve is specified for each farm in the dairy and horticulture industries, with large farms consuming more water at most prices, but all farms converging on similar maximum price points. For the broadacre industry, where estimation is not conducted on a per hectare basis, changes in farm size shift the water demand curve but do not change the slope. On broadacre (rice) farms, water demand actually decreases for farms with very large areas (which suggests that these farms commit only a small proportion of their total land area to irrigation). Overall the demand curves demonstrate the large effects that the assumed functional forms have on farm responses. Consideration of alternate function forms remains a subject for future research (see conclusions).

**Figure 5 Water demand curves for quartiles of farm size, by industry**

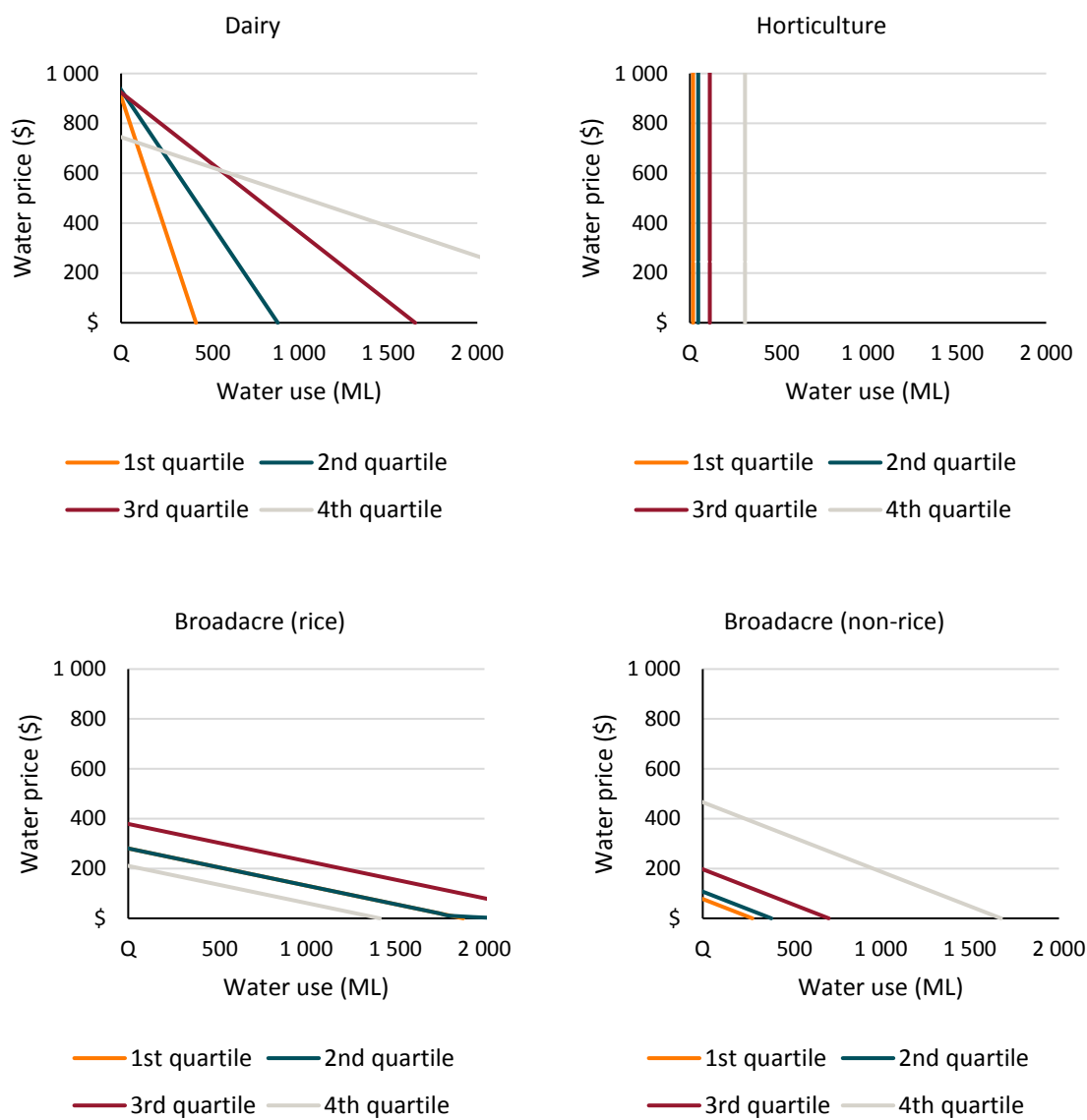

The water demand curves are calculated by taking the parameters of the linear water demand functions for each industry estimated in the farm production model, fixing all other variables at their mean values. The curves above give water demand in the Murrumbidgee region for dairy, Murrumbidgee for broadacre (rice), Vic Murray for horticulture and all regions for broadacre (non-rice).

## Validation results

$R^2$ statistics (which measure the variation in each dependent variable captured by the explanatory variables included in the model) are presented in appendix D. The models explain 53 to 74 per cent of the variation in inputs and outputs in dairy; 37 to 67 per cent in the broadacre (rice), 43 to 69 per cent in the broadacre (non-rice) and 38 to 83 per cent in the horticulture industry.

R[2] values are generally lower for input equations, particularly for labour demand. This can be at least partially explained by the large number of zero values for these variables. This is explored in more detail in appendix D, where the full R-squared values are presented, along with other model validation results including measures of monotonicity and convexity.

## Actual and predicted values

The model can also be assessed by its ability to replicate annual historical variation. The figures below compare actual and model predicted median values of total cost, total revenue and profit in each year. The model does a reasonable job of matching annual trends in farm costs, however in some cases revenue and in turn profit predictions differ significantly from actuals. In particular, the model over estimates revenue and profit on dairy and rice farms in some years.

**Figure 6 Average actual and predicted values, Dairy industry, 2006-07 to 2014-15**

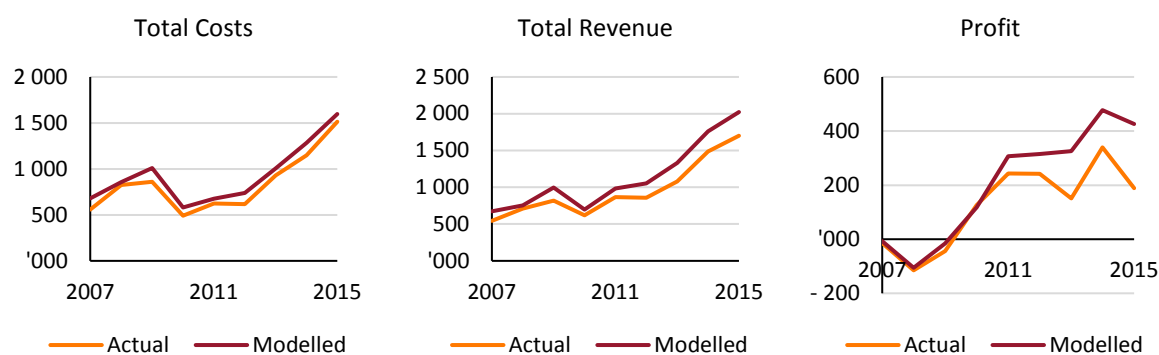

**Figure 7 Average actual and predicted values, Broadacre (rice) industry, 2006-07 to 2014-15**

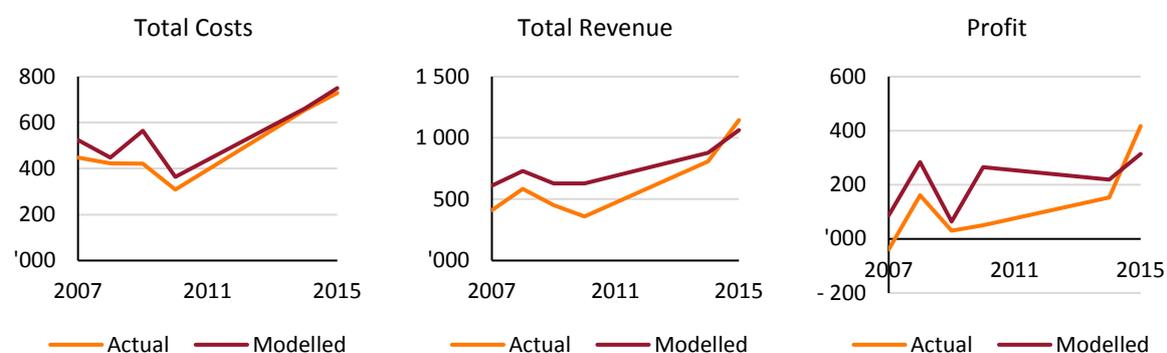

**Figure 8 Average actual and predicted values, Broadacre (non-rice) industry, 2006-07 to 2014-15**

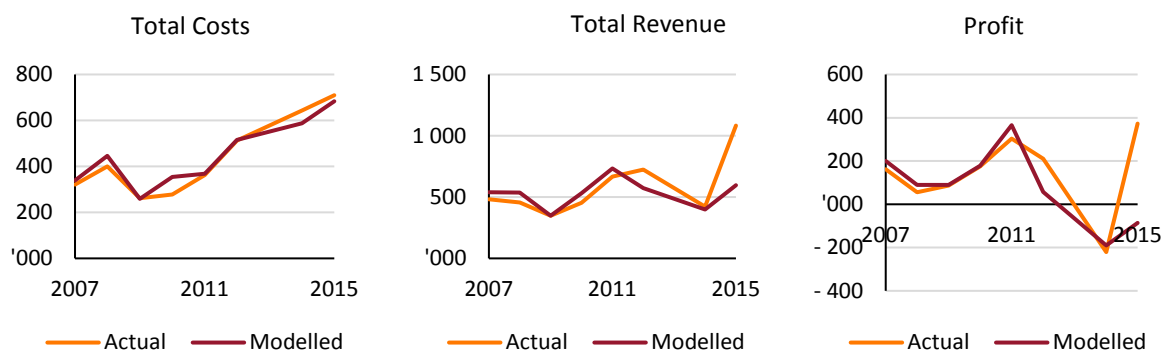

**Figure 9 Average actual and predicted values, Horticulture industry, 2006-07 to 2014-15**

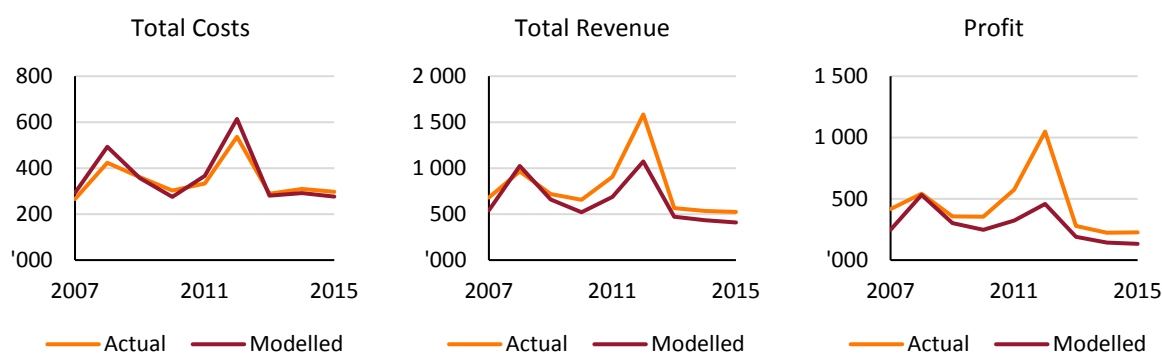

# Simulation model

The estimated input demand and output supply functions for each farm type are combined to form a micro-simulation model. Given scenario assumptions on fixed inputs, input and output prices and other control variables the model then simulates for each farm: input use and output supply. The changes to inputs and outputs are then aggregated to estimate responses to total costs, total revenue and profit. In the model, total profit is defined as the difference between total cost and total revenue, including the opportunity cost of water. As a result the model assumes that farmers pay for water use at the market price, rather than for the cost of delivering water alone. Profit also excludes any additional farm revenue streams outside of the variables included in the model as well as the effect of stocks in livestock or crops.

The simulation model produces results using the 2014-15 sample farms as a baseline. The model baseline scenario is defined by the model predictions for 2014-15 given the prices, climate conditions and other variables that applied during that year. Alternative scenarios are then constructed by varying one or more of the exogenous factors such as water prices. Appendix A also provides a more detailed description of the microsimulation model.

# 4 Results

Using the parameter estimates from the farm production model, ABARES used the microsimulation model to assess the economic impacts of a hypothetical 30 per cent increase in water allocation prices in the southern Murray-Darling Basin. 2014-15 predicted values are used for the baseline, and the estimated changes represent how an increase in water prices could have affected production and profit in 2014-15. Results presented in this section are the weighted average farm level results (unless otherwise specified). There are 276 farms in the model baseline (50 dairy farms, 49 broadacre (rice) farms, 34 broadacre (non-rice) farms and 143 horticulture farms).

The 30 per cent increase in allocation price scenario takes the 2014-15 water prices for each region included in the model and increases these values by 30 per cent. This is illustrated in figure 10 below, with the red bars showing the water price under the new scenario.

**Figure 10 Water price shock, 30 per cent increase on mean 2014-15 water prices for each region**

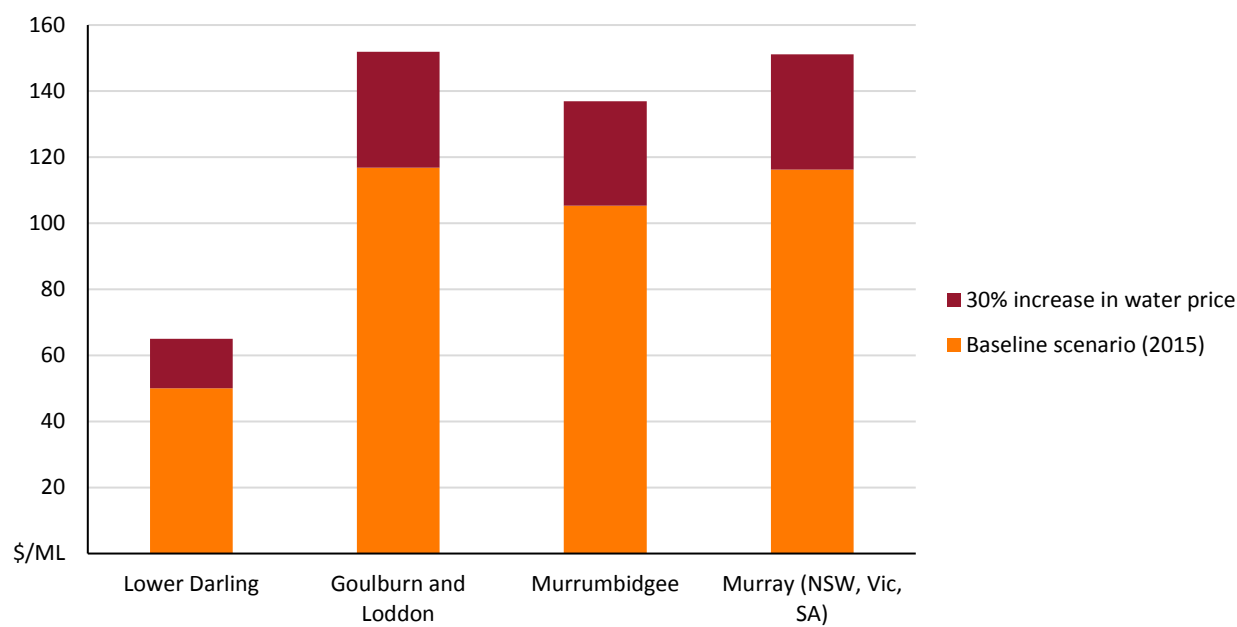

## Dairy farms

### Output and input quantities

Figure 11 shows weighted average farm-level estimates of changes in input and output quantities associated with a 30 per cent increase in allocation prices. The results suggest that dairy farmers will respond to the water price increase by reducing water use by 6.7 per cent and substituting towards fodder (up 2.5 per cent). Milk production is estimated to decrease by 1.1 per cent, while the model also shows a small increase in dairy cattle sales (up 1.1 per cent).

**Figure 11 Percentage change in input and output quantities at the farm level, dairy industry**

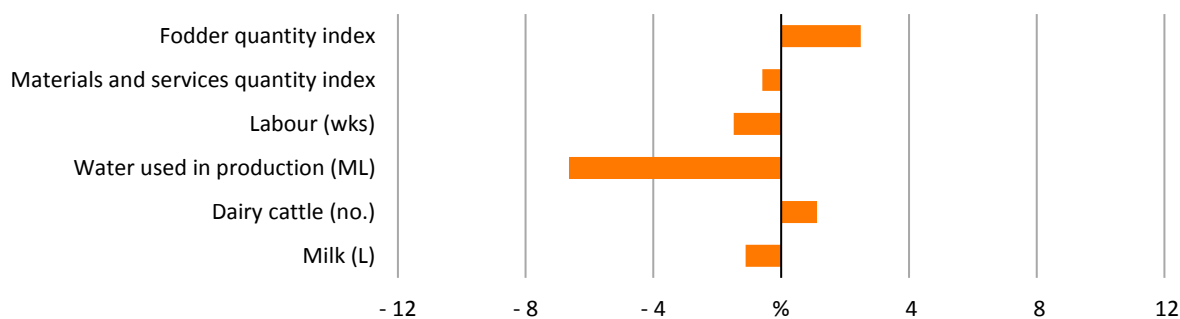

## Revenue, cost and profit

Table 5 shows how the increase in water prices affect farm level costs, revenue and profit. Note that the baseline estimated revenue and profit levels from the model for 2014-15 are above actual observed values (as shown in Figure 6).

The results indicate that a 30 per cent increase in allocation prices could lead to a $51,103 (14.4 per cent) reduction in farm profit, with this loss driven by a $28,926 (21.3 per cent) increase in water costs, and an increase in fodder expenditure (up $12,223). The increase in costs is only slightly offset by a decrease in labour costs (down $1,959). Despite a slight increase in dairy cattle sales, revenue decreases by $15,939 (0.9 per cent).

**Table 5 Percentage and level changes for farm costs, total revenue and total profit (weighted average)**

| Variables | Baseline (2014-15) | 30% increase in water price | Level change | Percentage change |
|---|---|---|---|---|
| Labour costs ($) | 131,553 | 129,593 | -1,959 | -1.5 |
| Fodder Costs ($) | 491,125 | 503,348 | 12,223 | 2.5 |
| materials and services costs ($) | 680,886 | 676,861 | -4,025 | -0.6 |
| Water costs ($) | 135,499 | 164,426 | 28,926 | 21.3 |
| Total cost ($) | 1,439,063 | 1,474,228 | 35,165 | 2.4 |
| Total revenue ($) | 1,793,893 | 1,777,954 | -15,939 | -0.9 |
| Profit ($) | 354,830 | 303,727 | -51,103 | -14.4 |

Note: Results are weighted averages, calculated at the farm level. Level and percentage change are calculated as the difference between the weighted average baseline and scenario values.

# Broadacre (rice) farms

## Output and input quantities

Figure 12 shows weighted average farm-level estimates of changes in input and output quantities in the broadacre (rice) industry due to a 30 per cent increase in allocation price. As with dairy, note that the baseline revenue and profit figures for rice farms are above observed values for 2014-15.

The results show that production declines in all outputs and inputs except for livestock. Rice production and other broadacre crops decrease on average by 9 and 9.9 per cent respectively. By contrast, livestock production increases by 2.2 per cent. This might reflect some expansion in dryland grazing activities as farmers reduce the area they use for irrigated crops.

The results show that a 30 per cent increase in allocation price leads to a more substantial reduction in water use on broadacre (rice) farms (down by 9.5 per cent) than on dairy or horticultural farms. The results also show a reduction in demand for labour (down 10.2 per cent).

**Figure 12 Percentage change in input and output quantities at the farm level, broadacre (rice) industry**

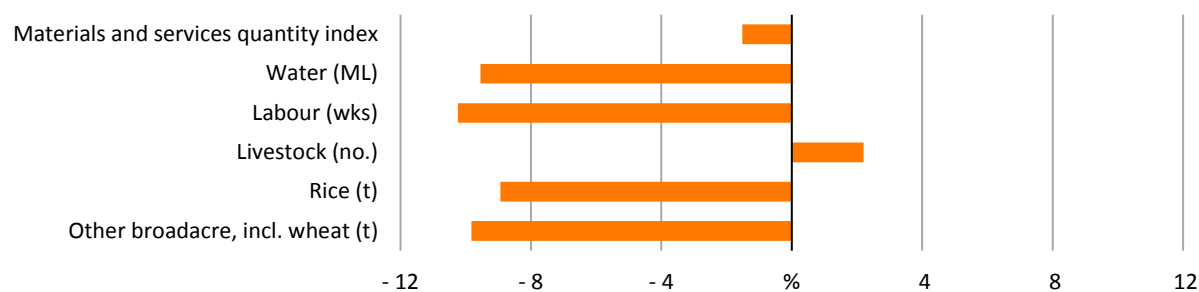

## Revenue, cost and profit

Table 6 shows changes in total revenue, total cost and total profit. The results show that a 30 per cent increase in allocation price leads to a $74,270 (25 per cent) reduction in profit, with increased costs and lower revenue contributing to the decline. Total costs increase by $18,558 (2.8 per cent), largely due to the increase in water costs. The $28,581 (17.5 per cent) increase in water costs is offset to some extent by a $3,040 (10.2 per cent) reduction in labour costs. Revenue falls by $55,712 (5.9 per cent), with lower revenue from crop production outweighing the increase in revenue from livestock production.

**Table 6 Percentage and level changes for farm costs, total revenue and total profit (weighted average)**

| Variable | Baseline (2014-15) | 30% increase in water price | Level change | Percentage change |
|---|---|---|---|---|
| Labour costs ($) | 29,687 | 26,646 | -3,040 | -10.2 |
| materials and services costs ($) | 460,435 | 453,452 | -6,983 | -1.5 |
| Water costs ($) | 163,365 | 191,946 | 28,581 | 17.5 |
| Total cost ($) | 653,486 | 672,044 | 18,558 | 2.8 |
| Total revenue ($) | 950,667 | 894,955 | -55,712 | -5.9 |
| Profit ($) | 297,181 | 222,911 | -74,270 | -25 |

Note: Results are weighted averages, calculated at the farm level. Level and percentage change are calculated as the difference between the weighted average baseline and scenario values.

# Broadacre (non-rice) farms

## Output and input quantities

Figure 13 shows weighted average farm-level estimates of changes in input and output quantities in the broadacre (non-rice) industry due to a 30 per cent increase in allocation price. It shows that water use declines while materials and services expenditure increases. Output production decreases to lesser extent than on rice farms with other broadacre crops decreasing by 3.1 per cent, while livestock production increases by 2.7 per cent. The largest reductions are in water use, which decreases by 6.6 per cent, slightly more than on broadacre (rice) farms.

**Figure 13 Percentage change in input and output quantities at the farm level, broadacre (non-rice) industry**

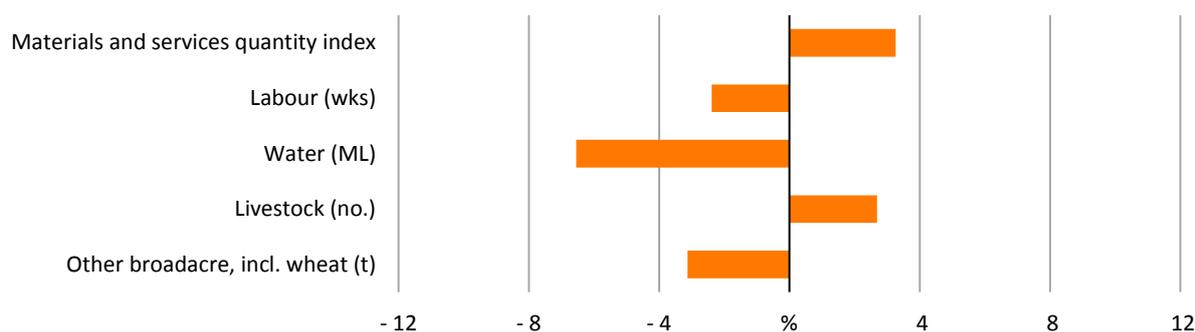

## Revenue, cost and profit

Table 7 shows changes in total revenue, total cost and total profit. The results show that a 30 per cent increase in allocation price leads to a $45,453 (35.9 per cent) reduction in profit, with increased costs contributing to the decline. Total costs increase by $41,763 (6.5 per cent), largely due to the increase in water costs of $27,185 (21.5 per cent) and materials and services costs of $15,512 (3.3 per cent). Total revenue also decreases by $3,609 (0.5 per cent), contributing to the reduction in profit.

**Table 7 Percentage and level changes for farm costs, total revenue and total profit (weighted average)**

| Variable | Baseline (2014-15) | 30% increase in water price | Level change | Percentage change |
|---|---|---|---|---|
| Labour costs ($) | 39,078 | 38,144 | -934 | -2.4 |
| materials and services costs ($) | 475,034 | 490,546 | 15,512 | 3.3 |
| Water costs ($) | 126,412 | 153,597 | 27,185 | 21.5 |
| Total cost ($) | 640,524 | 682,287 | 41,763 | 6.5 |
| Total revenue ($) | 767,283 | 763,593 | -3,690 | -0.5 |
| Profit ($) | 126,759 | 81,305 | -45,453 | -35.9 |

Note: Results are weighted averages, calculated at the farm level. Level and percentage change are calculated as the difference between the weighted average baseline and scenario values.

# Horticulture farms

## Output and input quantities

Figure 14 shows weighted average farm-level estimates of changes in input and output quantities on horticulture farms. While results in the previous sections are weighted average values for the full baseline sample in each industry, percentage changes in horticulture are calculated separately for each output type. For example, the reduction of citrus production of 0.08 per cent pertains only to farms who produce citrus. Changes to inputs are calculated in the same way as the above sections.

The results suggest that higher water prices lead to slight reductions in some outputs: stone fruit (0.8 per cent), table grapes (0.09 per cent), other horticultural (0.09 per cent), citrus (0.08 per cent) production and slight increases in others: vegetables (0.02 per cent), pome fruit (0.4 per cent) and wine grape production (0.5 per cent). Higher water prices also appear to result in some substitution away from materials and services towards hired labour, though it is unclear why this is the case.

**Figure 14 Percentage change in input and output quantities at the farm level, horticulture industry**

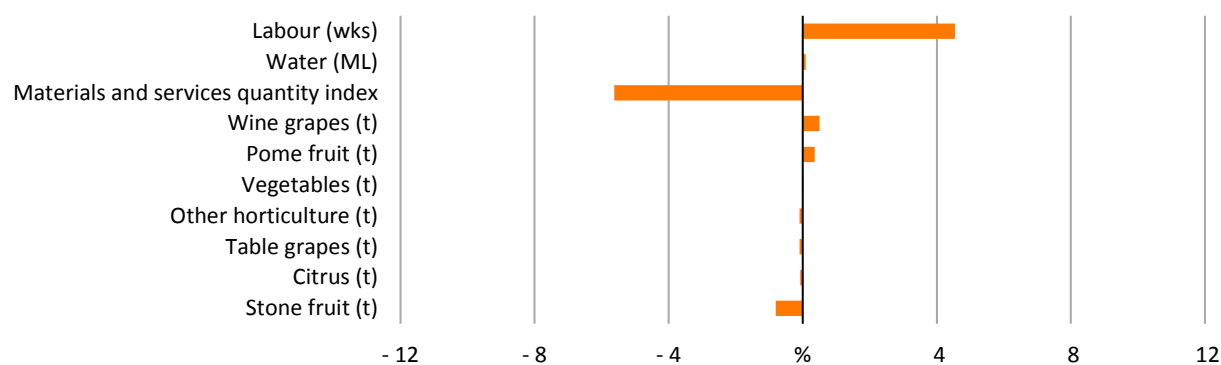

## Revenue, cost and profit

Table 8 gives changes to total revenue, total cost and total profit on horticulture farms. Overall, the results suggest that the effects on costs and profit are much less significant than for dairy and broadacre farms, with costs and profit effectively unchanged (costs decrease 0.5 per cent) (profit increases 0.5 per cent). Water costs increase by $4,653 (29.9 per cent). The change in revenue is even smaller (decreasing $92).

The water cost presented in table 7 is smaller than what would be expected from recent ABARES reports of Australia's horticultural industry (Ashton & Van Dijk, 2017). This is partially a reflection of the distribution of farm size in the horticultural industry baseline data. The 2014-15 sample data is skewed towards smaller horticulture producers (see the values for mean and median area operated in table 3).

**Table 8 Percentage and level changes for farm costs, total revenue and total profit (weighted average)**

| Variable | Baseline (2014-15) | 30% increase in water price | Level change | Percentage change |
|---|---|---|---|---|
| Labour costs ($) | 29,947 | 31,306 | 1,359 | 4.5 |
| materials and services costs ($) | 123,024 | 116,108 | -6,916 | -5.6 |
| Water costs ($) | 15,453 | 20,106 | 4,653 | 30.1 |
| Total cost ($) | 168,424 | 167,519 | -904 | -0.5 |
| Total revenue ($) | 325,939 | 325,847 | -92 | 0.0 |
| Profit ($) | 157,516 | 158,328 | 812 | 0.5 |

Note: Results are weighted averages, calculated at the farm level. Level and percentage change are calculated as the difference between the weighted average baseline and scenario values.

# Profit change across the MDB

The map below shows percentage changes in profit for all industries across the southern Murray-Darling Basin, with darker red showing a greater negative change in profit than lighter red. The map reflects the pattern for each industry shown above: larger increases in total cost and larger decreases in profit for the relatively lower value industries, broadacre (rice and non-rice) farms.

The map shows that greater profit loss is clustered around the Murrumbidgee, a rice growing region and the NSW Murray and northern Goulburn Broken regions, dominated by dairying. The Vic Murray and SA Murray regions, dominated by horticulture farming show only a small percentage decrease in profit. Despite water costs increasing substantially for horticultural farms, water costs remain a relatively small part of total costs and so total profit does not decrease by as much for horticulturalists as for other producers.

**Map 2 Estimated percentage change in profit for all industries**

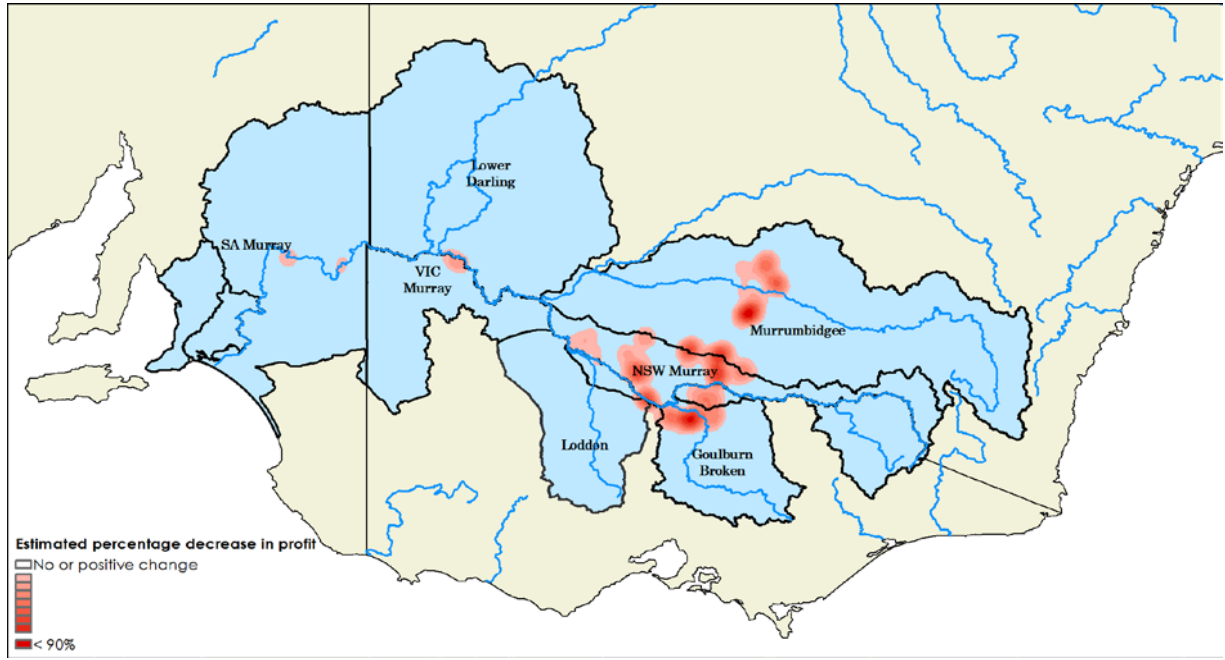

# 5 Conclusions

This report presented a micro-simulation model of irrigation farms in the sMDB, based on data from ABARES irrigation survey. The model can estimate farm level responses to exogenous shocks in the price of inputs, outputs or changes to climate. The model has a range of potential applications, such as examining the effects on irrigation farms of water policy reforms.

To demonstrate the model, we presented results for an arbitrary 30 per cent increase in the price of allocation water across the southern MDB. The results show the model is capable of representing the different types of responses expected on irrigation farms in different industries.

Broadacre farms are found to be the most responsive to increases in water prices (in terms of both changes in production and water demand). In contrast dairy farms, are more likely to maintain production in the short-run by substituting fodder and other inputs for irrigation water. Horticulture, farms are the least responsive to changes in water prices given their need to maintain permanent plantings. Horticulture farms tend to maintain water use and production in the face of water price increases, by incurring additional water costs. However, this has limited effect on profitability given that water costs area a relatively small percentage of total costs on horticulture farms. In contrast, water price increases have larger negative effects on profitability for dairy and rice farms.

While the responses of farms to changes in water prices broadly conform to expectations, the effect of other exogenous variables (i.e., input / output prices) are in some cases insignificant or outside of prior expectations. Further, while the model presents generally plausible average results, the predictions for individual farms are less realistic given the model's limited flexibility to account for farm heterogeneity (due to the assumed linear functional form).

A third limitation is that the model assumes irrigation farms purchase all water inputs at market value (i.e., buy allocations on the water market). While this assumption is consistent with economic theory (as farmers face the opportunity cost of water) there is also value in measuring actual farm cash income which will depend importantly on each farm's water entitlement holdings.

In light of these limitations there remain several options for further refinement of the model:

- More flexible functional forms could be employed, either parametric, semi-parametric (e.g., locally weighted least squares) or non-parametric. This might also involve relaxing the currently imposed profit maximisation constraints. Such an approach is dependent on having sufficient sample sizes, but could become more feasible over time as additional years of farm survey data become available.

- Another option is to supplement ABARES farm survey data, with farm level data from the ABS agricultural census and surveys. This may be possible in future through ABS-ABARES data sharing agreements. While ABS data is less detailed, it could for example be used to estimate farm level water demand functions, to a higher level of precision and resolution than possible with ABARES data.

- A water entitlement and allocation trade component could be added to the model. This version of the model would specify the entitlement holdings for each farm, and in–turn water allocations available to them under the baseline scenario. The model would then determine each farms allocation trade position and their net trade costs / proceeds.

# Appendix A: Model Structure

This section explains the econometric theories and calculations underlying the farm production and microsimulation models used to generate the results in this report.

## Farm production model

The farm production model used in this study is based on the traditional assumption of multiple-output profit maximisation (Chambers, 1988, Varian, 1992). In this study, ABARES has used the normalised quadratic (NQ) form of the indirect profit function. This flexible function is a second-order Taylor's approximation to the underlying production function that does not impose as many restrictions on production technology as other flexible functional forms such as Cobb-Douglas or CES (Shumway, 1983). It allows for global convexity to be imposed without losing flexibility, and is self-dual in that under certain conditions the production function and the normalised quadratic profit function contain the same information about the underlying production technology.

Drawing on Moschini (1988), we use the restricted NQ per hectare profit function in order to account for the differences in production responses by farm size:

$$\pi^*(p, z) = \sum_{i=1}^{n+m-1} a_i \times p_i + \sum_{f=1}^{k} b_f \times z_f + \frac{1}{2} \sum_{i=1}^{n+m-1} \sum_{j=1}^{n+m-1} c_{ij} \times p_i \times p_j + \frac{1}{2} \sum_{l=1}^{k} \sum_{f=1}^{k} d_{lf} \times z_l \times z_f$$
$$+ \sum_{i=1}^{n+m-1} \sum_{f=1}^{k} \propto_{if} p_i z_f \qquad (1)$$

where

$\pi^*$ - the restricted profit normalised by the price of materials and services $P_0$

$$\pi^* = {(\Pi/L)} \big/ P_0$$

$p$ – vector of output and input prices normalised by the price of materials and services

$p_i = P_i/P_0$;

$z$ – vector of $k$ fixed inputs;

$y$ - vector of $n$ outputs;

$x$ - vector of $(m-1)$ variable inputs

$a, b, c, d, \alpha$ - vectors of parameters to be estimated

Applying the Shephard-Hotelling's lemma, the following input demand and output supply functions are derived as the first derivative of the restricted profit function with respect to the input and output prices — so-called the netput prices.

The output supply function for output product i (i = 1 … n) is

$$y_i = \frac{\partial \Pi}{\partial p_i} = a_i + \sum_{j=1}^{n+m-1} c_{ij} \times p_j + \sum_{f=1}^{k} \propto_{if} \times z_f \qquad (2)$$

The negative input demand function for input h ($h = (n + 1, \ldots, n + m - 1)$) is

$$-x_h = \frac{\partial \Pi}{\partial p_h} = a_h + \sum_{j=1}^{n+m-1} c_{hj} \times p_j + \sum_{f=1}^{k} \propto_{hf} \times z_f \qquad (3)$$

Based on production theory, the estimated profit function must satisfy four conditions: linear homogeneity, symmetry, monotonicity and convexity. Homogeneity is imposed in equations (1) (2) and (3) but cannot be tested. The symmetricity condition requires that the second derivatives of the profit function with respect to netput prices and fixed inputs are independent of the order of derivation. That means, $c_{ij} = c_{ji}$ and $d_{lf} = d_{fl}$. For monotonicity, input demand and output supply estimates must be positive. For convexity, the Hessian matrix of output and negative input derivatives with respect to their prices must be positive semi-definite.

Following Shumway (1983), ABARES estimated the system of output supply and input demand equations specified in equations (2) and (3) with the imposition of homogeneity and symmetricity. The combination of homogeneity and symmetricity makes it possible to identify many parameters of the profit function and materials and services equations. Specifically, the estimation identifies vectors of parameters — $a, c, \alpha$ — which are used to calculate changes in farm-level and aggregate economic performance in response to changes in water prices.

Following Huffman and Evenson (1989), the coefficients on the netput prices in the materials and services equation can be derived based on the estimated coefficients in the system.

The negative input demand for the normalised input can be expressed as follows:

$$\pi^*(p, z) = \sum_{i=1}^{n} y_i \times p_i - \sum_{h=1}^{m-1} x_h \times p_h - x_m \quad (4)$$

Combining equations (1), (2) and (3) into (4), we get:

$$-x_m = a_m + \sum_{f=1}^{k} b_f \times z_f - \frac{1}{2} \sum_{i=1}^{n+m-1} \sum_{j=1}^{n+m-1} c_{ij} \times p_i \times p_j + \frac{1}{2} \sum_{l=1}^{k} \sum_{f=1}^{k} d_{lf} \times z_l \times z_f$$

$$+ \sum_{i=1}^{n+m-1} \sum_{f=1}^{k} \propto_{if} p_i z_f \qquad (5)$$

Marginal effect of the normalised input price to the output:

$$c_{im} = \frac{\partial y_i}{\partial P_m} = \sum_{j=1}^{n+m-1} \frac{\partial y_i}{\partial p_j} \times \frac{\partial p_j}{\partial P_m} = \sum_{j=1}^{n+m-1} c_{ij} \times \frac{\partial p_j}{\partial P_m} = \sum_{j=1}^{n+m-1} c_{ij} \times \frac{\partial (\frac{P_j}{P_m})}{\partial P_m}$$

$$= \sum_{j=1}^{n+m-1} c_{ij} \times \frac{(-P_j)}{(P_m)^2} = -\frac{1}{P_m} \sum_{j=1}^{n+m-1} c_{ij} \times p_j \qquad (6)$$

Marginal effect of the normalised input price to the input:

$$c_{hm} = \frac{\partial x_h}{\partial P_m} = \sum_{j=1}^{n+m-1} \frac{\partial x_h}{\partial p_j} \times \frac{\partial p_j}{\partial P_m} = -\sum_{j=1}^{n+m-1} c_{hj} \times \frac{\partial p_j}{\partial P_m} = -\sum_{j=1}^{n+m-1} c_{hj} \times \frac{\partial \left(\frac{P_j}{P_m}\right)}{\partial P_m}$$

$$= -\sum_{j=1}^{n+m-1} c_{hj} \times \frac{(-P_j)}{(P_m)^2} = \frac{1}{P_m} \sum_{j=1}^{n+m-1} c_{hj} \times p_j \qquad (7)$$

Marginal effect of other netput prices on the demand for the normalised input:

$$c_{mi} = \frac{\partial x_m}{\partial p_i} = \sum_{j=1}^{n+m-1} \frac{\partial y_j}{\partial p_i} \times p_j = \sum_{j=1}^{n+m-1} c_{ij} \times p_j \qquad (8)$$

The own-price marginal effect on the normalised input is:

$$c_{mm} = \frac{\partial x_m}{\partial P_m} = \sum_{i=1}^{n+m-1} \frac{\partial x_m}{\partial p_i} \times \frac{\partial p_i}{\partial P_m} = \sum_{j=1}^{n+m-1} \frac{\partial x_m}{\partial p_i} \times \frac{\partial (P_i/P_m)}{\partial P_m} = \sum_{j=1}^{n+m-1} \frac{\partial x_m}{\partial p_i} \times \left(-\frac{P_i}{(P_m)^2}\right)$$

$$= -\frac{1}{P_m} \sum_{j=1}^{n+m-1} \frac{\partial x_m}{\partial p_i} \times p_i = -\frac{1}{P_m} \sum_{i=1}^{n+m-1} \sum_{j=1}^{n+m-1} c_{ij} \times p_i \qquad (9)$$

The elasticity of output to the netput price:

$$\varepsilon_{ij} = c_{ij} \frac{p_j}{y_i} \qquad (10)$$

The elasticity of input to the netput price

$$\varepsilon_{hj} = c_{hj} \frac{p_j}{y_h} \qquad (11)$$

The elasticity of output and input to the normalised price $P_m$

$$\varepsilon_{im} = -\frac{1}{y_i} \sum_{j=1}^{n+m-1} c_{ij} \times p_j \qquad (12)$$

$$\varepsilon_{hm} = \frac{1}{x_h} \sum_{j=1}^{n+m-1} c_{hj} \times p_j \qquad (13)$$

The elasticity of the normalised input to the price of the netput:

$$\varepsilon_{mi} = \frac{p_i}{x_0} \sum_{j=1}^{n+m-1} c_{ij} \times p_j \qquad (14)$$

The own-price elasticity of the normalised input:

$$\varepsilon_{mm} = -\frac{1}{x_m} \sum_{i=1}^{n+m-1} \sum_{j=1}^{n+m-1} c_{ij} \times p_i \times p_j \quad (15)$$

# Microsimulation model

The estimated parameters for the input and output functions for each industry are used to simulate the effects of a 30 per cent increase in water allocation prices using farm-level survey data. This involves comparing results from the price increase scenario (s1) with the baseline scenario (s0). 2014-15 water prices are assumed in s0, while s1 assumes 2014-15 prices are 30 per cent higher). The farm-level estimates are then aggregated using survey weights to obtain aggregate changes at the industry level.

## Changes in farm-level economic performance in response to changes in water price

**Changes in output**

*Level change in output:*

$$\Delta y_i = y_{i,1} - y_{i,0} = c_{iw}\Delta p_w = c_{iw}(p_{w,0} - p_{w,1}) \quad (16)$$

$p_w$ is the normalised price of water

$c_{iw}$ is the marginal effect of water price on output or negative input i

*Output level under the policy change scenario:*

$$y_{i,1} = \Delta y_i + y_{i,0} \quad (17)$$

*Percentage change in output:*

$$\%\Delta y_i = \Delta y_i * 100/y_{i,1}$$

**Changes in input**

*Level change in input*

$$\Delta x_h = -c_{hw}\Delta p_w \quad (h = n+1,..,n+m-1) \quad (18)$$

*Level change in the normalised input (materials and services)*

$$\Delta x_{m+n} = \frac{\sum_{i=1}^{n} c_{iw}p_i - \sum_{i=1}^{n+m-1} c_{hw}p_h}{P_{n+m}} = \frac{\sum_{i=1}^{n+m-1} c_{iw}p_i}{P_0} \quad (19)$$

$P_0$ is the price index of materials and services

*Input level under the policy change scenario:*

$$x_{h,1} = \Delta x_h + x_{h,0} \quad (20)$$

*Percentage change in input:*

$$\%\Delta x_h = \Delta x_h * 100/x_{h,1} \quad (21)$$

**Profit change**

Changes in inputs and outputs due to a change in water price also lead to a change in total profit. ABARES defines total profit as the difference between total revenue and total cost, including the implicit cost of water. As a result, the analysis assumes that farmers pay for water use at the market price, rather than just for the cost of delivering water.

*Level change in total profit:*

$$
\begin{aligned}
\Delta\pi = \pi_1 - \pi_0 &= \left( \sum_{i=1}^{n} y_{i,1}\, p_{i,0} - \sum_{h=n+1}^{n+m-1} x_{h,1}\, p_{h,0} - x_{w,1} p_{w,1} \right) \\
&\quad - \left( \sum_{i=1}^{n} y_{i,0}\, p_{i,0} - \sum_{h=n+1}^{n+m-1} x_{h,0}\, p_{h,0} - x_{w,0} p_{w,0} \right) \\
&= \sum_{i=1}^{n} \Delta y_i\, p_{i,0} - \sum_{h=n+1}^{n+m-1} \Delta x_h\, p_{h,0} + x_{w,0} p_{w,0} - x_{w,1} p_{w,1} \\
&= \sum_{i=1}^{n} \Delta y_i\, p_{i,0} - \sum_{h=n+1}^{n+m-1} \Delta x_h\, p_{h,0} - x_{w,1}(p_{w,1} - p_{w,0}) - (x_{w,1} - x_{w,0}) p_{w,0} \\
&= \sum_{i=1}^{n} \Delta y_i\, p_{i,0} - \sum_{h=n+1}^{n+m-1} \Delta x_h\, p_{h,0} - x_{w,1} \Delta p_w - \Delta x_w p_{w,0} \qquad (22)
\end{aligned}
$$

*Profit under the policy change:*

$$\pi_1 = \Delta\pi + \pi_0 \qquad (23)$$

*Percentage change in profit:*

$$\%\Delta\pi = \Delta\pi * 100 / \pi_1 \qquad (24)$$

# Changes in aggregate economic performance in response to changes in water price

This section details how changes in input, output and profit can be aggregated from farm-level estimates to create industry level estimates.

**Aggregate changes in output/input quantities, profit:**

***Aggregate change in output/input***

*Level change in aggregate output:*

$$\Delta Q = \sum_{i=1}^{n} \Delta q_i \qquad (25)$$

$\Delta q_i$ is farm i output/input change

n is the number of farms in the region/the whole sample

*Total aggregate output/input under the policy change scenario*

$$Q_1 = \sum_{i=1}^{n} q_{i,1} \qquad (26)$$

$q_{i,1}$ is farm's i output/input under the policy change scenario

*Percentage change in aggregate output/input*

$$\%\Delta Q = \Delta Q * 100/Q_1 \tag{27}$$

**Aggregate change in total profit:**

*Level change in profit:*

$$\Delta\Pi = \sum_{i=1}^{n} \Delta\pi_i \tag{28}$$

$\Delta\Pi$ is the aggregate change in output/input at region level

$\Delta\pi_i$ is farm's i output/input change

*Total output/input under the policy change scenario:*

$$\Pi_1 = \sum_{i=1}^{n} \pi_{i,1} \tag{29}$$

$\pi_{i,1}$ is farm's i output/input change under the policy change scenario

*Percentage change in aggregate profit:*

$$\%\Delta\Pi = \Delta\Pi * 100/\Pi_1 \tag{30}$$

# Appendix B: Parameters

The farm microsimulation model outputs mean marginal effects for outputs and inputs in each industry. These give the average change in the use of an input or the production of an output in response to a change in the price of a given output or input.

On a per hectare basis, marginal effects are the same for all farms within an industry but are multiplied by farm size to give each farm a unique response in the dairy and horticulture industries. Tables 9 – 12 below, provide the mean marginal effects after accounting for farm size for each input and output in each industry.

**Table 9 Estimated parameters for dairy farms**

| Marginal effect on | With respect to the price of | | | | | |
|---|---|---|---|---|---|---|
| | Milk ($/L) | Dairy cattle ($/number) | Labour ($/weeks) | fodder (index) | Water ($/ML) | Materials and Services (index) |
| Milk (L) | **-211,732** | **525** | **-30** | **-648,556** | **-529** | **692,563** |
| se | (1278000) | (229) | (96) | (354317) | (269) | (400063) |
| pvalue | 0.87 | 0.02 | 0.76 | 0.07 | 0.05 | 0.08 |
| Dairy cattle (number) | **525** | **0** | **0** | **-173** | **0** | **-130** |
| se | (229) | (0) | (0) | (69) | (0) | (92) |
| pvalue | 0.02 | 0.40 | 0.28 | 0.01 | 0.23 | 0.16 |
| Labour (weeks) | **30** | **0** | **0** | **-83** | **0** | **-110** |
| se | (96) | (0) | (0) | (29) | (0) | (46) |
| pvalue | 0.76 | 0.28 | 0.32 | 0.00 | 0.34 | 0.02 |
| Fodder (index) | 648,556 | 173 | -83 | -428,940 | 179 | -86,324 |
| se | (354317) | (69) | (29) | (132524) | (83) | (106260) |
| pvalue | 0.07 | 0.01 | 0.00 | 0.00 | 0.03 | 0.42 |
| Water (ML) | 529 | 0 | 0 | 179 | -1 | 59 |
| se | (269) | (0) | (0) | (83) | (0) | (85) |
| pvalue | 0.05 | 0.23 | 0.34 | 0.03 | 0.00 | 0.49 |
| Materials and Services (index) | 692,563 | -130 | -110 | -86,324 | 59 | -64,251 |
| se | (400063) | (92) | (46) | (106260) | (85) | (182511) |
| pvalue | 0.08 | 0.16 | 0.02 | 0.42 | 0.49 | 0.73 |

Asymptotic standard errors are reported in parentheses; p value in italics.

**Table 10 Estimated parameters for broadacre (rice) farms**

| Marginal effect on | With respect to the price of | | | | | |
|---|---|---|---|---|---|---|
| | Rice ($/tonne) | Wheat ($/tonne) | Livestock ($/number) | Labour ($/weeks) | Water ($/ML) | Materials and Services (index) |
| Rice (tonne) | **-0.33** | **0.07** | **-0.99** | **-0.10** | **-1.83** | **1,245.00** |
| se | (0.26) | (0.21) | (0.32) | (0.03) | (0.36) | (198.90) |
| pvalue | 0.20 | 0.73 | 0.00 | 0.00 | 0.00 | 0.00 |
| Other broadacre, incl. wheat  (tonne) | **0.07** | **-0.30** | **-1.22** | **-0.06** | **-1.97** | **1,348.00** |
| se | (0.21) | (0.33) | (0.28) | (0.03) | (0.38) | (214.20) |
| pvalue | 0.73 | 0.37 | 0.00 | 0.02 | 0.00 | 0.00 |
| Livestock (number) | **-0.99** | **-1.22** | **1.02** | **0.10** | **2.69** | **-171.00** |
| se | (0.32) | (0.28) | (0.95) | (0.13) | (0.54) | (432.10) |
| pvalue | 0.00 | 0.00 | 0.28 | 0.45 | 0.00 | 0.69 |
| Labour (weeks) | **0.10** | **0.06** | **-0.10** | **0.04** | **-0.09** | **69.30** |
| se | (0.03) | (0.03) | (0.13) | (0.08) | (0.05) | (76.69) |
| pvalue | 0.00 | 0.02 | 0.45 | 0.64 | 0.05 | 0.37 |
| Water (ML) | **1.83** | **1.97** | **-2.69** | **-0.09** | **-6.59** | **-1,190.00** |
| se | (0.36) | (0.38) | (0.54) | (0.05) | (0.73) | (300.90) |
| pvalue | 0.00 | 0.00 | 0.00 | 0.05 | 0.00 | 0.00 |
| Materials and Services (index) | **1,245.00** | **1,348.00** | **-171.00** | **69.30** | **1,190.00** | **-1,108,000.00** |
| se | (198.90) | (214.20) | (432.10) | (76.69) | (300.90) | (268705.00) |
| pvalue | 0.00 | 0.00 | 0.69 | 0.37 | 0.00 | 0.00 |

Note: Asymptotic standard errors are reported in parentheses; p value in italics.

**Table 11 Estimated parameters for broadacre (non-rice) farms**

| Marginal effect on | With respect to the price of | | | | |
|---|---|---|---|---|---|
| | Wheat ($/tonne) | Livestock ($/number) | Labour ($/weeks) | Water ($/ML) | Materials and Services (index) |
| Other broadacre, incl. wheat (tonne) | **-0.21** | **-0.16** | **0.00** | **-1.15** | **427.70** |
| se | (0.24) | (0.18) | (0.01) | (0.24) | (126.70) |
| pvalue | 0.39 | 0.36 | 0.71 | 0.00 | 0.00 |
| Livestock (number) | **-0.16** | **1.66** | **0.02** | **0.85** | **-849.50** |
| se | (0.18) | (0.93) | (0.06) | (0.38) | (419.50) |
| pvalue | 0.36 | 0.08 | 0.78 | 0.03 | 0.04 |
| Labour (weeks) | **0.00** | **-0.02** | **0.00** | **-0.03** | **-9.68** |
| se | (0.01) | (0.06) | (0.03) | (0.02) | (38.32) |
| pvalue | 0.71 | 0.78 | 0.96 | 0.24 | 0.80 |
| Water (ML) | **1.15** | **-0.85** | **-0.03** | **-3.59** | **-738.50** |
| se | (0.24) | (0.38) | (0.02) | (0.52) | (204.30) |
| pvalue | 0.00 | 0.03 | 0.24 | 0.00 | 0.00 |
| Materials and Services (index) | **427.70** | **-849.50** | **-9.68** | **-738.50** | **358,765.00** |
| se | (126.70) | (419.50) | (38.32) | (204.30) | (205608.00) |
| pvalue | 0.00 | 0.04 | 0.80 | 0.00 | 0.08 |

Note: Asymptotic standard errors are reported in parentheses; p value in italic

## Table 12 Estimated parameters for horticulture farms

| Marginal effect on | With respect to the price of | | | | | | | | | |
|---|---|---|---|---|---|---|---|---|---|---|
| | Pome ($/tonne) | Citrus ($/tonne) | Stone fruit ($/tonne) | Table grapes ($/tonne) | Wine grapes ($/tonne) | Vegetables ($/tonne) | Other Horticulture ($/tonne) | Labour ($/week) | Water ($/ML) | Materials and services (index) |
| Pome (tonne) | -0.02 | 0.02 | 0.01 | 0.00 | -0.02 | -0.01 | 0.00 | 0.03 | 0.02 | -24.45 |
| se | (0.01) | (0.03) | (0.01) | (0.00) | (0.01) | (0.00) | (0.00) | (0.02) | (0.01) | (32.92) |
| pvalue | 0.16 | 0.47 | 0.32 | 0.11 | 0.03 | 0.00 | 0.76 | 0.16 | 0.09 | 0.46 |
| Citrus (tonne) | 0.02 | 0.17 | -0.02 | -0.01 | -0.05 | 0.00 | 0.00 | 0.06 | -0.01 | -83.76 |
| se | (0.03) | (0.19) | (0.02) | (0.00) | (0.05) | (0.01) | (0.01) | (0.02) | (0.02) | (43.87) |
| pvalue | 0.47 | 0.36 | 0.33 | 0.20 | 0.25 | 0.80 | 0.84 | 0.01 | 0.78 | 0.06 |
| Stone fruit (tonne) | 0.01 | -0.02 | -0.01 | 0.00 | 0.01 | 0.00 | 0.00 | 0.00 | -0.02 | 21.06 |
| se | (0.01) | (0.02) | (0.01) | (0.00) | (0.01) | (0.00) | (0.00) | (0.01) | (0.01) | (22.55) |
| pvalue | 0.32 | 0.33 | 0.12 | 0.45 | 0.12 | 0.30 | 0.31 | 0.95 | 0.03 | 0.35 |
| Table grapes (tonne) | 0.00 | -0.01 | 0.00 | 0.00 | 0.00 | 0.00 | 0.00 | 0.00 | 0.00 | 9.36 |
| se | (0.00) | (0.00) | (0.00) | (0.00) | (0.00) | (0.00) | (0.00) | (0.00) | (0.00) | (6.59) |
| pvalue | 0.11 | 0.20 | 0.45 | 0.69 | 0.53 | 0.63 | 0.11 | 0.39 | 0.54 | 0.16 |
| Wine grapes (tonne) | -0.02 | -0.05 | 0.01 | 0.00 | -0.02 | -0.01 | 0.00 | -0.03 | 0.03 | 66.27 |
| se | (0.01) | (0.05) | (0.01) | (0.00) | (0.02) | (0.00) | (0.00) | (0.02) | (0.01) | (27.39) |
| pvalue | 0.03 | 0.25 | 0.12 | 0.53 | 0.29 | 0.00 | 0.67 | 0.11 | 0.02 | 0.02 |
| vegetables (tonne) | -0.01 | 0.00 | 0.00 | 0.00 | -0.01 | 0.01 | 0.00 | -0.02 | 0.00 | 9.39 |
| se | (0.00) | (0.01) | (0.00) | (0.00) | (0.00) | (0.01) | (0.00) | (0.01) | (0.00) | (16.81) |
| pvalue | 0.00 | 0.80 | 0.30 | 0.63 | 0.00 | 0.01 | 0.66 | 0.08 | 0.84 | 0.58 |
| Other horticulture (tonne) | 0.00 | 0.00 | 0.00 | 0.00 | 0.00 | 0.00 | 0.00 | 0.00 | 0.00 | 5.81 |
| se | (0.00) | (0.01) | (0.00) | (0.00) | (0.00) | (0.00) | (0.00) | (0.00) | (0.00) | (8.77) |
| pvalue | 0.76 | 0.84 | 0.31 | 0.11 | 0.67 | 0.66 | 0.67 | 0.31 | 0.76 | 0.51 |
| Labour (weeks) | -0.03 | -0.06 | 0.00 | 0.00 | 0.03 | 0.02 | 0.00 | 0.03 | 0.07 | -85.90 |
| se | (0.02) | (0.02) | (0.01) | (0.00) | (0.02) | (0.01) | (0.00) | (0.08) | (0.02) | (125.10) |
| pvalue | 0.16 | 0.01 | 0.95 | 0.39 | 0.11 | 0.08 | 0.31 | 0.75 | 0.00 | 0.49 |
| Water (ML) | -0.02 | 0.01 | 0.02 | 0.00 | -0.03 | 0.00 | 0.00 | 0.07 | 0.00 | -132.60 |
| se | (0.01) | (0.02) | (0.01) | (0.00) | (0.01) | (0.00) | (0.00) | (0.02) | (0.02) | (39.50) |
| pvalue | 0.09 | 0.78 | 0.03 | 0.54 | 0.02 | 0.84 | 0.76 | 0.00 | 0.93 | 0.00 |
| Materials and services (index) | 24.45 | 83.76 | -21.06 | -9.36 | -66.27 | -9.39 | -5.81 | -85.90 | -132.60 | 264,384.00 |
| se | (32.92) | (43.87) | (22.55) | (6.59) | (27.39) | (16.81) | (8.77) | (125.10) | (39.50) | (220441.00) |
| pvalue | 0.46 | 0.06 | 0.35 | 0.16 | 0.02 | 0.58 | 0.51 | 0.49 | 0.00 | 0.23 |

Note: Asymptotic standard errors are reported in parentheses; p value in it

# Appendix C: Elasticities

## Price elasticities

The estimated parameters can be used to derive price elasticities for inputs and outputs, which measure the percentage change in an input or output in response to a percentage change in price. Elasticity values for each of the industries are shown below in tables 13 - 16.

In most cases the responses of farms to changes in prices do not conform to prior expectations. Firstly, not all of the own price elasticities for outputs are positive. According to economic theory, when the price of an output goes up, we should expect farms to produce more of that output. Secondly, many of the own price elasticities for inputs are also positive. According to theory, when an input price goes up farms should use less of that input, not more.

## Cross price elasticities

Cross price elasticities also do not always align with expectations. This is particularly the case where statistically significant relationships could not be found. For example, the estimates suggest that an increase in the price of water will lead to small increases in pome fruit and wine grapes on horticulture farms. Such effects could in theory be the result of input substitution on horticulture farms with multiple activities but this is unlikely to be the case in reality.

Some of the cross-price elasticities are more reassuring. In the dairy industry, there is substitution between water and fodder (when the water price increases dairy farms demand more fodder and less water). For outputs in the broadacre industries, there is generally a substitution effect between crops and livestock, with livestock production increasing as crop prices decrease. There is a similar substitution effect between milk and dairy cattle on dairy farms.

**Table 13 Price elasticities of input and output quantities in the dairy industry**

| Elasticities of | With respect to the price of | | | | | |
|---|---|---|---|---|---|---|
| | Milk ($/L) | Dairy cattle ($/number) | Labour ($/weeks) | Fodder (price index) | Water ($/ML) | Materials and Services (index) |
| Milk (L) | **-0.08** | **0.16** | **-0.04** | **-0.49** | **-0.07** | **0.55** |
| se | (0.52) | (0.07) | (0.12) | (0.28) | (0.04) | (0.34) |
| pvalue | 0.87 | 0.02 | 0.76 | 0.07 | 0.05 | 0.11 |
| Dairy cattle (number) | **1.36** | **-0.17** | **0.28** | **-0.87** | **0.06** | **-0.68** |
| se | (0.60) | (0.21) | (0.26) | (0.34) | (0.05) | (0.50) |
| pvalue | 0.02 | 0.40 | 0.28 | 0.01 | 0.23 | 0.19 |
| Labour (weeks) | **0.34** | **-0.33** | **-0.71** | **-1.89** | **-0.05** | **2.65** |
| se | (1.11) | (0.31) | (0.72) | (0.67) | (0.06) | (1.13) |
| pvalue | 0.76 | 0.28 | 0.32 | 0.00 | 0.34 | 0.02 |
| Fodder (index) | **2.06** | **0.44** | **-0.81** | **-2.62** | **0.16** | **0.59** |
| se | (1.16) | (0.18) | (0.29) | (0.83) | (0.08) | (0.74) |
| pvalue | 0.07 | 0.01 | 0.00 | 0.00 | 0.03 | 0.44 |
| Water (ML) | **0.71** | **-0.06** | **-0.08** | **0.47** | **-0.49** | **-0.32** |
| se | (0.36) | (0.05) | (0.08) | (0.21) | (0.05) | (0.25) |
| pvalue | 0.05 | 0.23 | 0.34 | 0.03 | 0.00 | 0.00 |
| Materials and Services (index) | **-1.11** | **0.17** | **0.56** | **0.28** | **-0.03** | **0.00** |
| se | (0.67) | (0.12) | (0.24) | (0.35) | (0.05) | (0.64) |
| pvalue | 0.11 | 0.19 | 0.02 | 0.44 | 0.00 | 0.76 |

Note: Asymptotic standard errors are reported in parentheses; p value in italics. The elasticities are evaluated at the mean values of prices and quantity.

**Table 14 Price elasticities of input and output quantities in the broadacre (rice) industry**

| Elasticities of | With respect to the price of | | | | | |
|---|---|---|---|---|---|---|
| | Rice ($/tonne) | Wheat ($/tonne) | Livestock ($/number) | Labour ($/weeks) | Water ($/ML) | Materials and Services (index) |
| Rice (tonne) | **-0.43** | **0.11** | **-0.99** | **-0.35** | **-1.51** | **3.17** |
| se | (0.33) | (0.32) | (0.32) | (0.10) | (0.30) | (0.51) |
| pvalue | 0.20 | 0.73 | 0.00 | 0.00 | 0.00 | 0.00 |
| Other broadacre. Incl. wheat (tonne) | **0.04** | **-0.21** | **-0.54** | **-0.09** | **-0.72** | **1.51** |
| se | (0.12) | (0.23) | (0.12) | (0.04) | (0.14) | (0.24) |
| pvalue | 0.73 | 0.37 | 0.00 | 0.02 | 0.00 | 0.00 |
| Livestock (number) | **-0.57** | **-0.84** | **0.46** | **0.16** | **0.98** | **-0.19** |
| se | (0.18) | (0.20) | (0.42) | (0.21) | (0.20) | (0.49) |
| pvalue | 0.00 | 0.00 | 0.28 | 0.45 | 0.00 | 0.69 |
| Labour (weeks) | **1.71** | **1.20** | **-1.38** | **1.91** | **-1.05** | **-2.41** |
| se | (0.47) | (0.53) | (1.81) | (4.05) | (0.55) | (2.66) |
| pvalue | 0.00 | 0.02 | 0.45 | 0.64 | 0.05 | 0.37 |
| Water (ML) | **0.97** | **1.27** | **-1.11** | **-0.14** | **-2.23** | **1.24** |
| se | (0.19) | (0.25) | (0.22) | (0.07) | (0.25) | (0.31) |
| pvalue | 0.00 | 0.00 | 0.00 | 0.05 | 0.00 | 0.00 |
| Materials and Services (index) | **-1.20** | **-1.57** | **0.13** | **-0.19** | **0.73** | **2.09** |
| se | (0.19) | (0.25) | (0.32) | (0.21) | (0.18) | (0.51) |
| pvalue | 0.00 | 0.00 | 0.69 | 0.37 | 0.00 | 0.00 |

Note: Asymptotic standard errors are reported in parentheses; p value in italics. The elasticities are evaluated at the mean values of prices and quantity.

**Table 15 Price elasticities of input and output quantities in the broadacre (non-rice) industry**

| Elasticities of | With respect to the price of | | | | |
|---|---|---|---|---|---|
| | Wheat ($/tonne) | Livestock ($/number) | Labour ($/weeks) | Water ($/ML) | Materials and Services (index) |
| Other broadacre. Incl. wheat (tonne) | **-0.06** | **-0.05** | **0.00** | **-0.19** | **0.31** |
| se | (0.07) | (0.05) | (0.01) | (0.04) | (0.09) |
| pvalue | 0.39 | 0.36 | 0.71 | 0.00 | 0.00 |
| Livestock (number) | **-0.07** | **0.68** | **0.03** | **0.19** | **-0.83** |
| se | (0.07) | (0.38) | (0.09) | (0.09) | (0.41) |
| pvalue | 0.36 | 0.08 | 0.78 | 0.03 | 0.04 |
| Labour (weeks) | **0.06** | **-0.27** | **0.09** | **-0.22** | **0.34** |
| se | (0.15) | (0.96) | (1.73) | (0.19) | (1.36) |
| pvalue | 0.71 | 0.78 | 0.96 | 0.24 | 0.80 |
| Water (ML) | **0.70** | **-0.51** | **-0.06** | **-1.18** | **1.05** |
| se | (0.15) | (0.23) | (0.05) | (0.17) | (0.29) |
| pvalue | 0.00 | 0.03 | 0.24 | 0.00 | 0.00 |
| Materials and Services (index) | **-0.37** | **0.72** | **0.03** | **0.35** | **-0.72** |
| se | (0.11) | (0.35) | (0.11) | (0.10) | (0.42) |
| pvalue | 0.00 | 0.04 | 0.80 | 0.00 | 0.08 |

Note: Asymptotic standard errors are reported in parentheses; p value in italics. The elasticities are evaluated at the mean values of prices and quantity.

## Table 16 Price elasticities of input and output quantities in the horticulture industry

| Elasticities of | With respect to the price of | | | | | | | | | |
|---|---|---|---|---|---|---|---|---|---|---|
| | Pome ($/tonne) | Citrus ($/tonne) | Stone fruit ($/tonne) | Table grapes ($/tonne) | Wine grapes ($/tonne) | Vegetables ($/tonne) | Other Horticulture ($/tonne) | Labour ($/week) | Water ($/ML) | Materials and services (index) |
| Pome (tonne) | -1.14 | 0.68 | 3.70 | 1.90 | -0.27 | -1.18 | 0.30 | 0.80 | 0.08 | -0.15 |
| se | (0.81) | (0.94) | (3.77) | (1.19) | (0.12) | (0.36) | (0.95) | (0.57) | (0.04) | (0.23) |
| pvalue | 0.16 | 0.47 | 0.32 | 0.11 | 0.03 | 0.00 | 0.76 | 0.16 | 0.09 | 0.48 |
| Citrus (tonne) | 1.29 | 5.86 | -8.35 | -3.68 | -0.60 | 0.16 | -0.64 | 1.96 | -0.02 | -0.46 |
| se | (1.80) | (6.37) | (8.66) | (2.90) | (0.47) | (0.64) | (3.21) | (0.74) | (0.07) | (0.23) |
| pvalue | 0.47 | 0.36 | 0.33 | 0.20 | 0.25 | 0.80 | 0.84 | 0.01 | 0.78 | 0.06 |
| Stone fruit (tonne) | 0.58 | -0.69 | -5.52 | -0.75 | 0.16 | 0.26 | 0.71 | 0.03 | -0.08 | 0.57 |
| se | (0.59) | (0.71) | (3.59) | (1.00) | (0.09) | (0.25) | (0.69) | (0.40) | (0.04) | (0.59) |
| pvalue | 0.32 | 0.33 | 0.12 | 0.45 | 0.12 | 0.30 | 0.31 | 0.95 | 0.03 | 0.29 |
| Table grapes (tonne) | 0.20 | -0.21 | -0.51 | 0.15 | 0.02 | 0.04 | -0.39 | -0.10 | -0.01 | 0.22 |
| se | (0.13) | (0.16) | (0.68) | (0.38) | (0.03) | (0.07) | (0.24) | (0.12) | (0.01) | (0.16) |
| pvalue | 0.11 | 0.20 | 0.45 | 0.69 | 0.53 | 0.63 | 0.11 | 0.39 | 0.54 | 0.13 |
| Wine grapes (tonne) | -1.54 | -1.82 | 5.89 | 1.00 | -0.23 | -0.96 | 0.55 | -0.76 | 0.11 | 0.36 |
| se | (0.73) | (1.57) | (3.78) | (1.59) | (0.20) | (0.33) | (1.28) | (0.48) | (0.05) | (0.15) |
| pvalue | 0.03 | 0.25 | 0.12 | 0.53 | 0.29 | 0.00 | 0.67 | 0.11 | 0.02 | 0.01 |
| vegetables (tonne) | -0.74 | 0.05 | 1.03 | 0.21 | -0.10 | 1.36 | -0.16 | -0.60 | 0.00 | 0.01 |
| se | (0.23) | (0.21) | (0.99) | (0.43) | (0.03) | (0.55) | (0.37) | (0.34) | (0.02) | (0.11) |
| pvalue | 0.00 | 0.80 | 0.30 | 0.63 | 0.00 | 0.01 | 0.66 | 0.08 | 0.84 | 0.47 |
| Other horticulture (tonne) | 0.05 | -0.05 | 0.73 | -0.59 | 0.02 | -0.04 | -0.14 | -0.15 | 0.00 | 0.19 |
| se | (0.16) | (0.27) | (0.72) | (0.37) | (0.03) | (0.10) | (0.33) | (0.15) | (0.01) | (0.26) |
| pvalue | 0.76 | 0.84 | 0.31 | 0.11 | 0.67 | 0.66 | 0.67 | 0.31 | 0.76 | 0.45 |
| Labour (weeks) | -1.67 | -2.17 | -0.35 | 1.95 | 0.28 | 1.99 | 1.95 | 0.76 | 0.25 | -3.80 |
| se | (1.19) | (0.81) | (5.38) | (2.26) | (0.16) | (1.14) | (1.91) | (2.45) | (0.08) | (6.03) |
| pvalue | 0.16 | 0.01 | 0.95 | 0.39 | 0.11 | 0.08 | 0.31 | 0.75 | 0.00 | 0.59 |
| Water (ML) | -1.40 | 0.20 | 9.25 | 0.98 | -0.35 | -0.10 | 0.33 | 2.27 | 0.01 | -1.75 |
| se | (0.83) | (0.71) | (4.36) | (1.60) | (0.14) | (0.49) | (1.06) | (0.72) | (0.08) | (0.52) |
| pvalue | 0.09 | 0.78 | 0.03 | 0.54 | 0.02 | 0.84 | 0.76 | 0.00 | 0.93 | 0.00 |
| Materials and services (index) | 0.28 | 0.47 | -0.71 | -0.21 | -0.50 | -0.18 | -0.26 | -1.42 | -0.26 | 3.01 |
| se | (0.39) | (0.25) | (0.76) | (0.16) | (0.22) | (0.44) | (0.42) | (2.28) | (0.08) | (2.76) |
| pvalue | 0.48 | 0.06 | 0.34 | 0.17 | 0.01 | 0.43 | 0.46 | 0.60 | 0.00 | 0.29 |

Note: Asymptotic standard errors are reported in parentheses; p value in italics. The elasticities are evaluated at the mean values of prices and quantity.

# Appendix D: Validation results

## R squared

Table 17 shows the regression R-squared (per hectare) and level R-squared values. As the broadacre industry regressions are not run on a per hectare basis, only the level R2 values are included. R-squared values provide an indication of the predictive ability of a model, with an R-squared value closer to 1 indicating that the model has greater predictive ability. As would be expected, R-squared values are generally higher for the level variables than the per hectare variables.

The model has lower predictive ability for equations with dependent variables that contain a large number of zero observations, for example labour use and rice production. In these cases the model struggles to accurately predict zero production and will generally over estimate.

**Table 17 Actual and level R squared values for each regression**

| Dependent Variable | Per hectare R2 | Level R2 |
|---|---|---|
| **Dairy** | | |
| Milk (L) | 0.5 | 0.7 |
| Dairy cattle (no.) | 0.4 | 0.5 |
| Fodder (index) | 0.3 | 0.5 |
| Labour (weeks) | 0.3 | 0.7 |
| Water (ML) | 0.4 | 0.6 |
| **Broadacre (Rice)** | | |
| Rice (tonne) | | 0.5 |
| Other broadacre, incl. wheat (tonne) | | 0.5 |
| Livestock (number) | | 0.7 |
| Labour (weeks) | | 0.4 |
| Water (ML) | | 0.5 |
| **Broadacre (Non-Rice)** | | |
| Other broadacre, incl. wheat (tonne) | | 0.6 |
| Livestock (number) | | 0.7 |
| Labour (weeks) | | 0.5 |
| Water (ML) | | 0.5 |
| **Horticulture** | | |
| Pome (tonne) | 0.7 | 0.8 |
| Citrus (tonne) | 0.8 | 0.7 |
| Stone fruit (tonne) | 0.7 | 0.8 |
| Table grapes (tonne) | 0.6 | 0.8 |
| Wine grapes (tonne) | 0.8 | 0.8 |
| vegetables (tonne) | 0.4 | 0.8 |
| Other horticulture (tonne) | 0.3 | 0.5 |
| Labour (weeks) | 0.0 | 0.4 |
| Water (ML) | 0.3 | 0.5 |

## Goodness of fit

Figures 15 to 18 diagrammatically illustrate the goodness of fit for each input and output equation by charting actual (Y) and predicted ($\hat{Y}$) values for each dependent variable in each industry. The actual values are the observed results from the MDBIS data while the predicted values are the results predicted by the model. Charting the two values gives an indication of the models predictive ability. The wider the dispersion of orange dots around the red 45 degree line, the less accurate the model is at predicting changes to a variable of interest.

The model generally has a reasonable goodness of fit for the dairy industry and most outputs in the horticulture and broadacre (rice and non-rice) industries. The model often struggles with inputs, especially hired labour and with livestock in the broadacre regressions, where the model tends to underestimate for larger farms.

**Figure 15 Actual against predicted values, dairy industry**

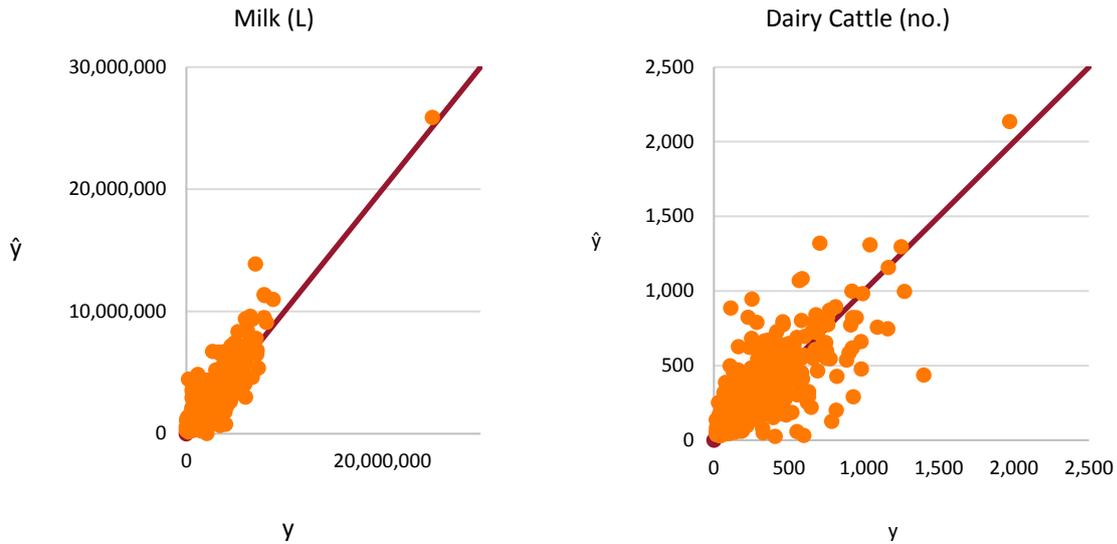

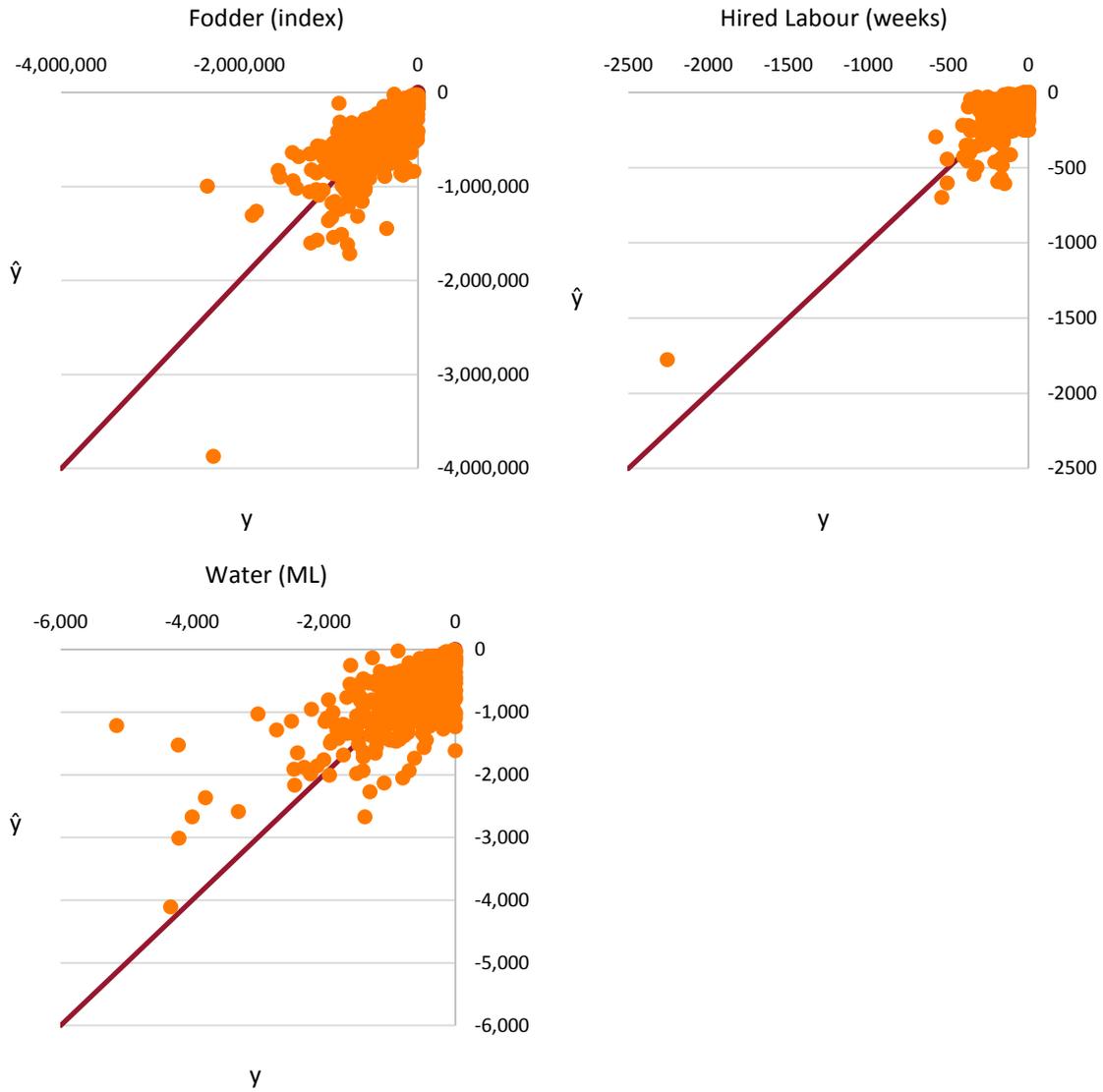

Figure 16 Actual against predicted values, broadacre (rice) industry

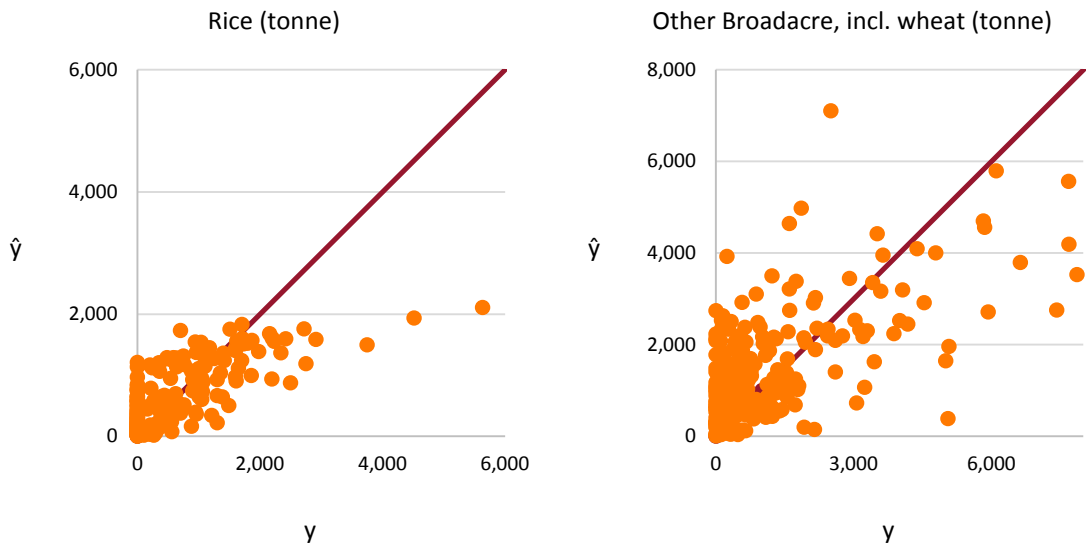

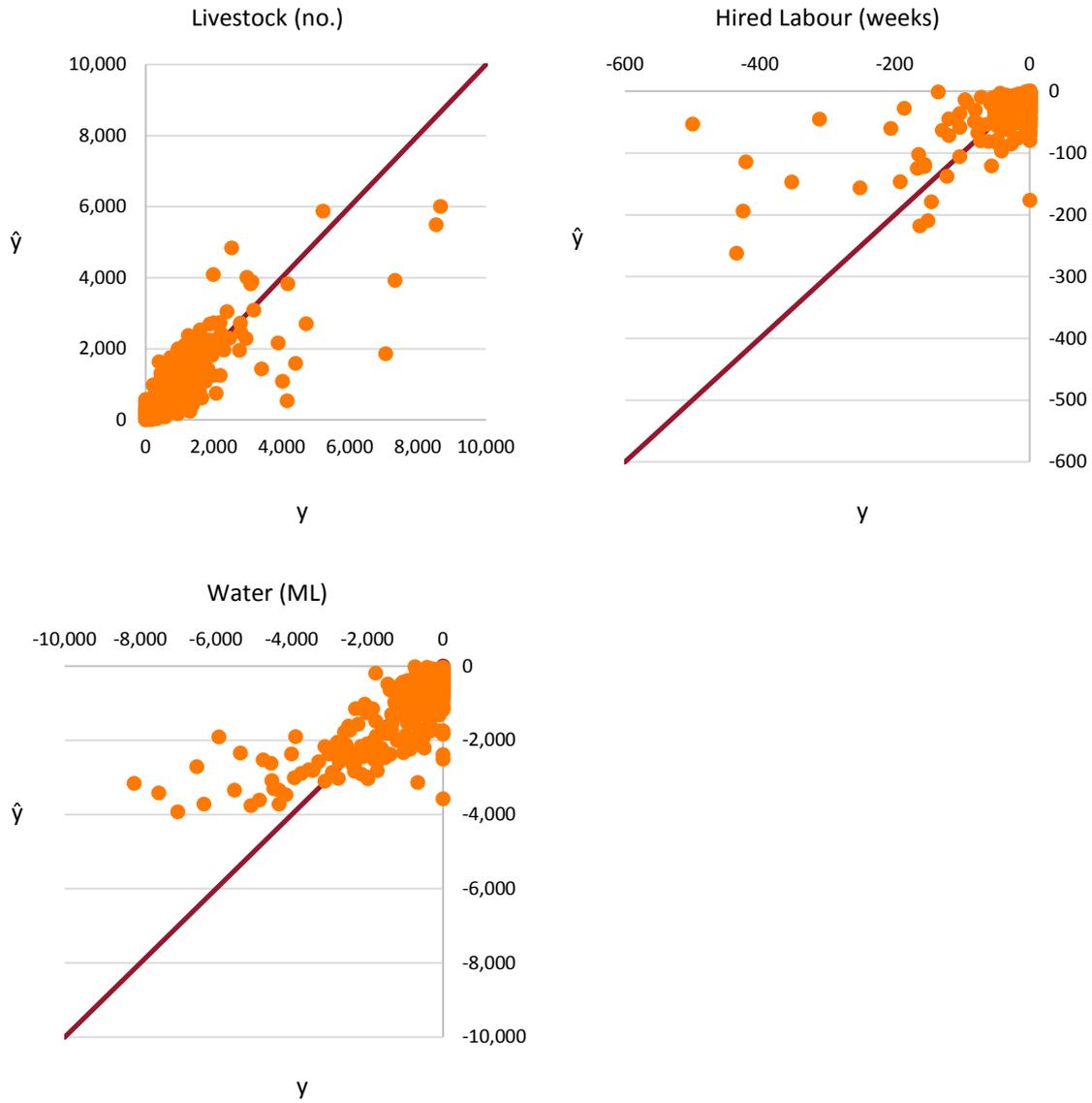

**Figure 17 Actual against predicted values, broadacre (non-rice) industry**

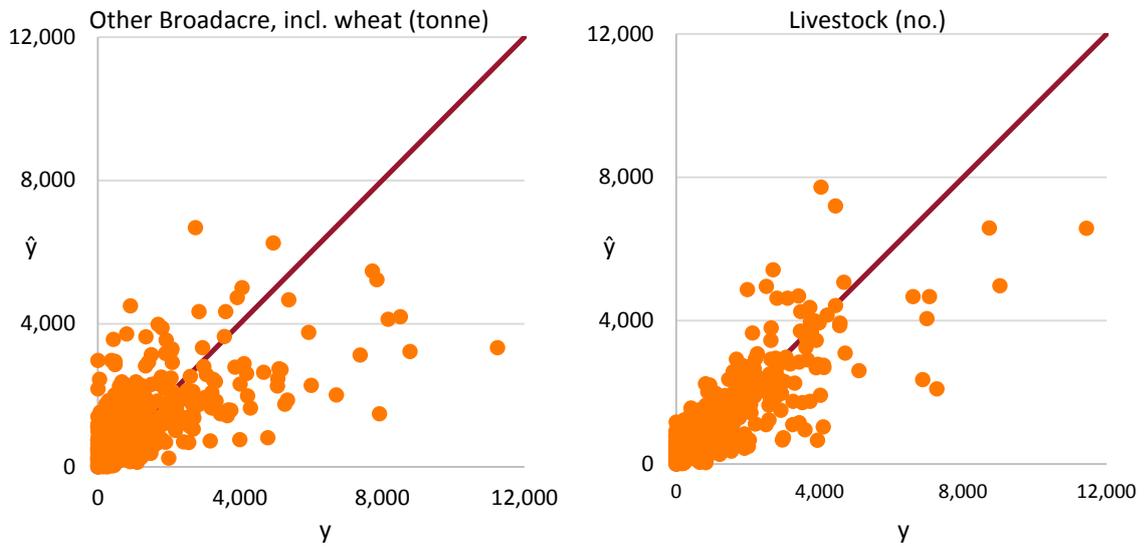

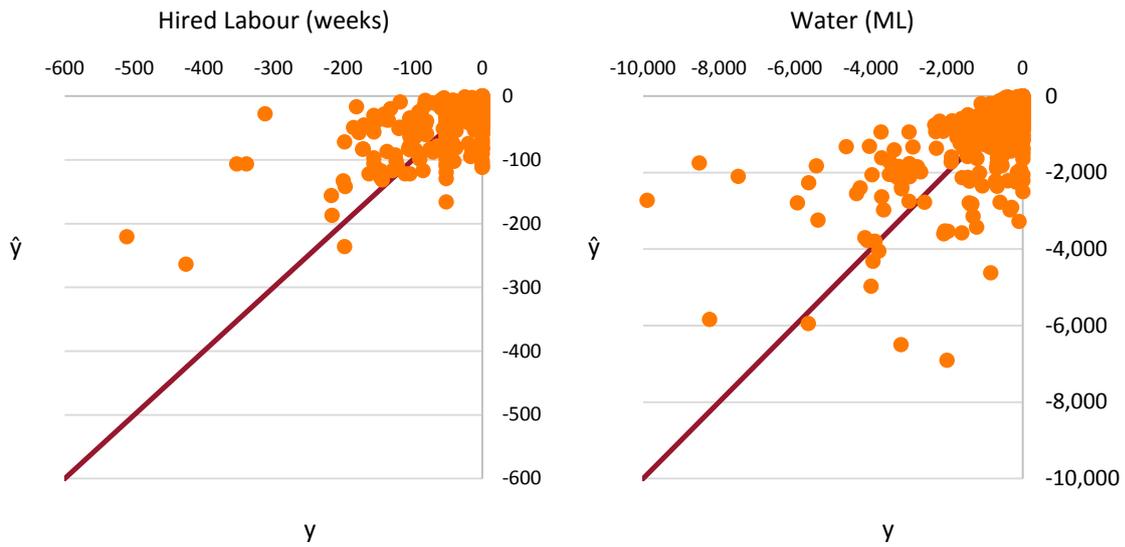

**Figure 18 Actual against predicted values, horticulture industry**

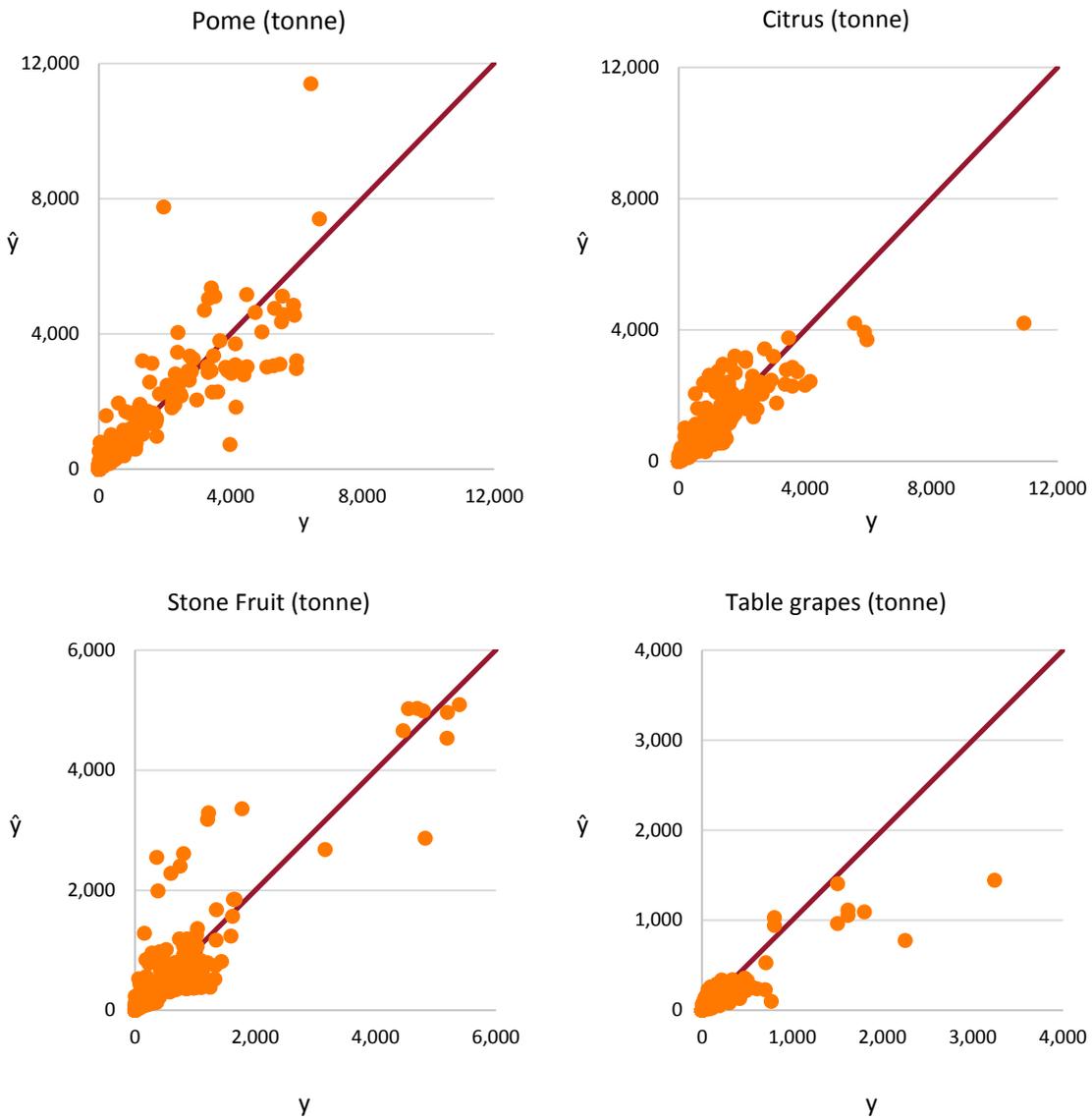

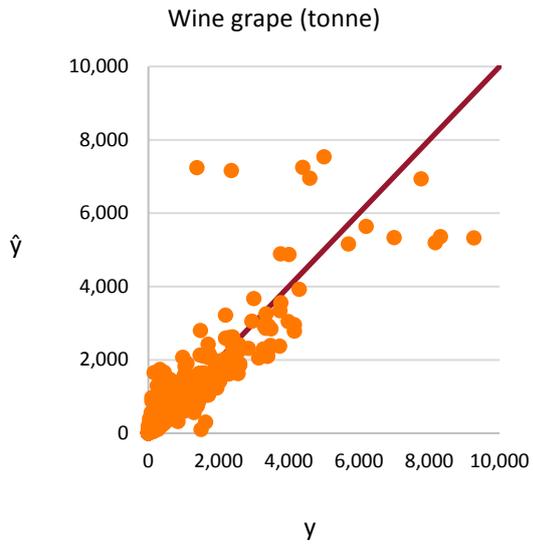

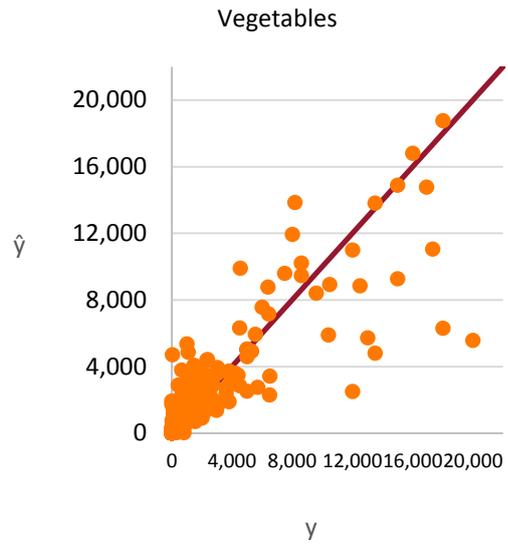

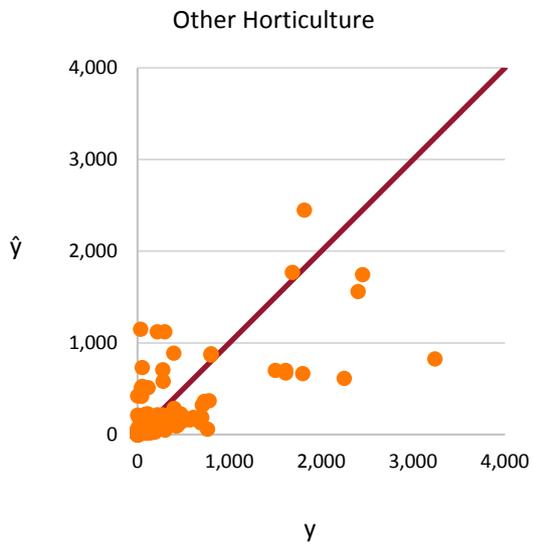

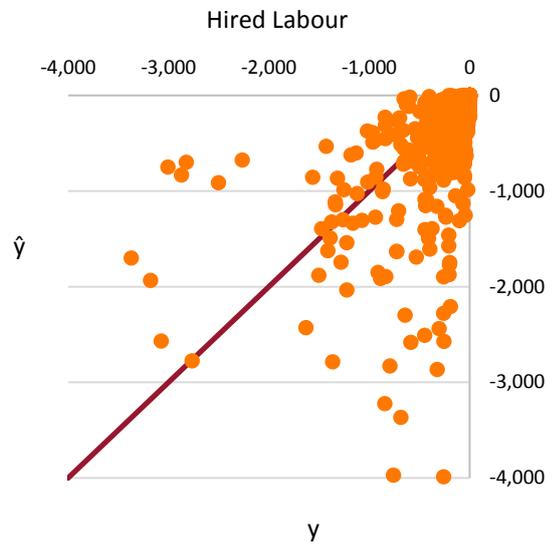

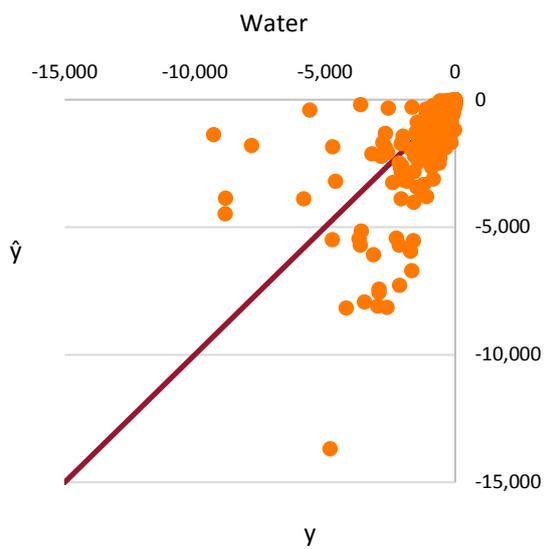

# Monotonicity

Monotonicity measures the degree to which the model satisfies the theoretical assumption that production responds in a consistent way to changes in prices. This is measured by the percentage of predicted values for each regression which have the expected sign – positive for outputs and negative for inputs. Monotonicity results are presented in table 18, below.

In most cases, regressions satisfy monotonicity conditions over 90 per cent of the time. Again, the horticulture industry performs most poorly, satisfying monotonicity conditions less than 60 per cent of the time for the pome, citrus and stone fruit regressions and less than 50 per cent of the time for the vegetable regression.

**Table 18 Monotonicity values for each regression**

| Dependent Variable | Monotonicity |
|---|---|
| **Dairy** | |
| Milk (L) | 1.0 |
| Dairy cattle (no.) | 1.0 |
| Fodder (index) | 0.9 |
| Labour (weeks) | 1.0 |
| Water (ML) | 1.0 |
| **Broadacre (Rice)** | |
| Rice (tonne) | 0.7 |
| Other broadacre, incl. wheat (tonne) | 0.7 |
| Livestock (number) | 0.9 |
| Labour (weeks) | 0.7 |
| Water (ML) | 0.8 |
| **Broadacre (Non-Rice)** | |
| Other broadacre, incl. wheat (tonne) | 0.8 |
| Livestock (number) | 0.9 |
| Labour (weeks) | 0.7 |
| Water (ML) | 0.8 |
| **Horticulture** | |
| Pome (tonne) | 0.5 |
| Citrus (tonne) | 0.5 |
| Stone fruit (tonne) | 0.5 |
| Table grapes (tonne) | 0.4 |
| Wine grapes (tonne) | 0.8 |
| vegetables  (tonne) | 0.4 |
| Other horticulture (tonne) | 0.8 |
| Labour (weeks) | 0.6 |
| Water (ML) | 1.0 |

# Convexity

Convexity refers to the extent that own-price elasticities have the expected sign – positive for outputs and negative for inputs. For outputs, this implies that as the price of the output increases, the quantity of the output produced will also increase. The opposite is true for inputs, with increases in price expected to lead to a decrease in use.

For most industries, the sign of own-price elasticities are contrary to expectations. Convexity is not satisfied for the pome fruit, stone fruit, wine grapes, other horticulture and labour regression in the horticultural industry, for other broadacre (including wheat), and labour regressions in the broadacre (non-rice) industry, for the rice and other broadacre regressions in the broadacre (rice) industry and for the milk and dairy cattle regressions in the dairy industry.

The overall convexity of inputs and outputs within an industry can be evaluated by looking at the Hessian matrix using Cholesky parameterisation (Lau 1976). Convexity conditions are satisfied if the Hessian matrix is positive semi-definite. None of the industries yield a positive semi-definite Hessian matrix. Though this appears poor, it is not entirely unexpected as convexity conditions are often not satisfied in applied econometric studies. Moreover, in almost all cases where convexity is not satisfied, the elasticities are not statistically significant. This implies that the model was not able to find a relationship between the quantity of the variable purchased and the price.